\documentclass[twocolumn,superscriptaddress,longbibliography,aps,prb,floatfix,
preprintnumbers]{revtex4-2}

\usepackage[normalem]{ulem}
\usepackage{graphicx}
\usepackage{bm}
\usepackage{color}
\usepackage{amsmath}
\usepackage{amssymb}
\usepackage{epstopdf}
\usepackage{lipsum}
\usepackage{gensymb}
\usepackage{threeparttable}
\usepackage{multirow}
\usepackage{scrextend}
\usepackage{xfrac}

\usepackage[urlcolor=blue,colorlinks=true,citecolor=blue,linkcolor=blue,pdfstartview={FitH},bookmarks=false]{hyperref}

\usepackage[dvipsnames]{xcolor}
\definecolor{mygreen}{rgb}{0.0, 0.6, 0.0}
\definecolor{pjorange}{rgb}{0.8, 0.3, 0.0}
\definecolor{jlblue}{rgb}{0.2, 0.5, 0.7}

\graphicspath{{fig/}{./fig/}{.}}

\begin{document}

\title{
Anisotropic electron--phonon coupling and chiral phonons \\
in van der Waals room temperature ferromagnet Fe$_{5}$GeTe$_{2}$
}

\author{Smrutiranjan~Mekap}
\affiliation {Department of Physics, Indian Institute of Technology Kharagpur, Kharagpur 721302, India} 

\author{Andrzej~Ptok}
\email{aptok@mmj.pl}
\affiliation{Institute of Nuclear Physics, Polish Academy of Sciences, ul. W. E. Radzikowskiego 152, 31-342 Krak\'{o}w, Poland}

\author{Jyoti~Saini}
\affiliation{School of Physical Sciences, Jawaharlal Nehru University, New Delhi-110067, India}

\author{Changgu~Lee}
\affiliation{SKKU Advanced Institute of Nanotechnology (SAINT), Sungkyunkwan University, Suwon 16419, Republic of Korea}
\affiliation{School of Mechanical Engineering, Sungkyunkwan University, Suwon 16419, Republic of Korea}

\author{Subhasis~Ghosh}
\email{subhasis.ghosh.jnu@gmail.com}
\affiliation{School of Physical Sciences, Jawaharlal Nehru University, New Delhi-110067, India}

\author{Anushree~Roy}
\email{anushree@phy.iitkgp.ac.in}
\affiliation {Department of Physics, Indian Institute of Technology Kharagpur, Kharagpur 721302, India}

\date{\today}

\begin{abstract}
The layered van der Waals Fe$_5$GeTe$_2$ (F5GT) compound exhibits room-temperature ferromagnetism, making it a promising candidate for technological applications.
In our study, combined temperature, wavelength, and polarization-dependent Raman measurements, along with {\it ab initio} calculations reveal important aspects of lattice dynamics and electron--phonon interactions.
The angle-resolved Raman intensity under linear polarization configurations exhibits a strong tilt in the laboratory coordinate system, indicating the existence of anisotropic electron--phonon coupling. 
The temperature evolution of this anisotropy is discussed by extracting the phase factor of the Raman tensor elements from the angle-resolved intensity measured at different temperatures, also uncovering a spin--orbit coupling--mediated electron--phonon response in F5GT.
The thermal evolution of electron--phonon coupling is also examined by measuring the temperature dependence of the Fano parameter of the asymmetric peak in the Raman spectra, while wavelength-dependent measurements establish the role of optical resonance in enhancing the anisotropic interaction.
Finally, the threefold rotational symmetry guarantees the existence of chiral phonons.
We present direct spectroscopic evidence for these chiral vibrational modes through cross-circularly polarized Raman measurements, complemented by theoretical calculations of phonon circular polarization. 
Together, these results identify F5GT as an ideal platform for investigating emergent couplings among lattice, electronic, and magnetic degrees of freedom and for advancing the understanding of chiral phonons in magnetic van der Waals materials.
\end{abstract}

\maketitle

\section{Introduction}
\label{sec.intro}

Recent technological progress in spintronics requires the discovery of materials that exhibit novel features~\cite{sierra2021,yang2021a,gish2024,zhao2025}.
One of the promising groups of materials is van der Waals (vdW) magnetic materials, which exhibit large spin--orbit coupling, high spin polarization, and a significant anomalous Hall angle~\cite{kim2018,yang2021,mi2023,kajale2024,zhang2025}.
Layered structures with weak vdW coupling enable simple fabrication of different types of heterostructures for potential technological applications~\cite{geim2013,novoselov2016,zhong2017,
wang2020,xu2024,zollner2025}.

One example of such promising vdW ferromagnetic compounds is the Fe$_{x}$GeTe$_{2}$ ($3 \leq x \leq 7$) (FGT) family~\cite{seo2020,liu2022,jiang2022,ren2023,adhikari2025}.
These family members exhibit nearly room temperature ferromagnetism~\cite{deiseroth2006,chen2013,may2016,stahl2018,may2019a,li2018,may2019b,zhang2020c}.
The realized crystal structure can be investigated as a combination of transition metal dichalcogenides and body-centered cubic Fe lattices~\cite{liu2022,seo2020}.
The main layout of Fe-Fe dumbbells forms multiple layers of Fe-rich slabs sandwiched between two Te layers, one above and one below, making it suitable for vdW stacking [e.g., see Fig.~\ref{fig.crystal}(a)]. 
Increasing the Fe sublayers is crucial for enhancing the pair exchange interaction and improving magnetic ordering~\cite{liu2022,jiang2022}.
For example, Fe$_{3}$GeTe$_{2}$ (F3GT) exhibits $T_\text{C} = 220$~K~\cite{deiseroth2006,chen2013}, and this can be enhanced to $270$~K for Fe$_{4}$GeTe$_{2}$ (F4GT)~\cite{monda2021,liu2022}.
However, observed magnetism is very sensitive to doping~\cite{tian2020,zhang2024} or the number of layers in the system~\cite{wang2025}.
In this paper, we focus on the Fe-rich Fe$_{5}$GeTe$_{2}$ (F5GT), which has the highest $T_\text{C} \simeq 260$--$310$~K among FGT family members~\cite{stahl2018,may2019a}. 
F5GT exhibits several phenomena that are interesting from the viewpoint of its possible electronic applications~\cite{lin.peng.2022,ren2023}. For example, due to the ferromagnetism of F5GT, the anomalous Hall effect was reported~\cite{may2019a,zhao2023,suzuki2023,moon2024,zhang2024,yu2025}, which confirms its possible applications in quantum devices. 

F5GT exhibits a strong connection between the magnetic ground states, local atomic order, and layer stacking~\cite{may2020}.
The role of structural disorder is observed during thermal cycling of magnetization from $350$ K to $2$ K and back to $350$ K~\cite{may2019a}. 
During the initial thermal cycle down to $2$~K, a first-order magneto-structural phase transition occurs at low temperatures $\sim 100$~K~\cite{may2019a,may2019b,chen2023}. 
Further, the existence of spin--orbit coupling~\cite{Bera2024} and the itinerant characteristic of the electron~\cite{yadav2024} result in an overall complex interplay between spin-lattice-electronic-orbital degrees of freedom in this system. 
Electron--phonon coupling is expected to play a central role in governing this intricate microscopic mechanism of such a phenomenon, as lattice vibrations can modulate spin-polarized electronic states and, in turn, be renormalized by magnetic order. 
Further, a study on chiral phonons is especially important in the magnetically ordered state, where their angular momentum can interact with spin polarization. Such lattice-spin coupling opens pathways to unconventional helicity-dependent dynamics and enriches the broader landscape of collective excitations in magnetic van der Waals materials.
Raman spectroscopy, as a sensitive probe of lattice dynamics in {\it quasi}-two-dimensional materials, may shed new light on uncovering such features.

We report several notable signatures of lattice
dynamics in F5GT: {\it (i)} Anisotropic electron--phonon coupling: 
In magnetic layered materials, anisotropic electron--phonon coupling has typically been observed under an externally applied magnetic field~\cite{mccreary2020,McCreary2020b,sun2025}.
In contrast, experimental evidence and \textit{ab initio} calculations presented in this article demonstrate optical resonance-induced electron-phonon coupling in F5GT without relying on an external magnetic field. Furthermore, we report the experimental evidence of the role of spin--orbit coupling, even in the paramagnetic phase, which is crucial for understanding the microscopic origins of the magneto-optic effects in F5GT, essential for its potential technological applications. {\it (ii)} Existence of chiral phonons: Crystals possessing threefold rotational symmetry and broken time-reversal symmetry can host chiral phonons, which are associated with circular atomic motion around their equilibrium positions ~\cite{wang2024}. We present direct experimental evidence for these modes in F5GT using cross-circularly polarized Raman measurements, supported by first-principles calculations of phonon circular polarization. Notably, these chiral vibrational states emerge intrinsically, without the need for an external magnetic field.

The paper is organized as follows.
The computational methodology is described in Sec.~\ref{sec.mainmethod}.
Our results and their discussion are presented in Sec.~\ref{sec.results}.
Finally, Sec.~\ref{sec.sum} summarizes the main findings and conclusions of this work.

\section{Methodology}
\label{sec.mainmethod}

\subsection{Experimental Details}

The single-crystal F5GT was grown by chemical vapor transport. 
The details of the sample growth parameters are available in~\cite{Tiwari2024}.
Crystallographic axes orientations of the crystal are available in Fig.~\ref{fig.orientation} in the Supplemental Material (SM)~\footnote{See Supplemental Material at [URL will be inserted by publisher] for additional experimental and theoretical results. This Supplemental Material includes Refs.~\cite{}}. 
The crystal was cleaved just before measurements.
Temperature-dependent magnetic measurements were carried out by applying a low magnetic field ($0.01$~T) as the sample was cooled from room temperature to low temperatures. 

Micro-Raman measurements were performed using a triple-stage monochromator Raman spectrometer, \mbox{T-64000} (Horiba, France). 
The spectrometer is equipped with a confocal microscope (Olympus, Japan) and a Peltier-cooled charge-coupled device (CCD) camera (Syncerity, USA). 
All spectra are recorded in the backscattering geometry using a $50 \times$ L objective lens. 
A $532$~nm Nd:YAG laser (Excelsior, USA) was used as the excitation source for all measurements. 
Furthermore, laser excitation wavelengths of $488.0$~nm ($2.54$~eV), $514.5$~nm ($2.41$~eV), and $568.2$~nm ($2.18$~eV) using a multiwavelength laser (Innova, 70C, Coherent) were employed for wavelength-dependent Raman measurements.
The sample stage THMS600 (Linkam, USA), equipped with a cryostat and a liquid nitrogen pump, is used to carry out temperature-dependent measurements. All measurements are carried out with 1 mW laser power on the sample. Checks were performed to rule out laser-induced heating effects by recording Raman spectra at different laser powers and monitoring any possible peak shifts.

For polarization-resolved Raman spectroscopy, a linear polarizer was placed in the incidence path, along with a half-wave plate (HWP) and an analyzer in the detection path. The HWP on the scattered path was rotated while the intensity of the scattered light was recorded at regular intervals of its polarization directions.
For helicity-resolved measurements, the laser is guided by a polarizer. 
In the incident light path, a half-wave plate (HWP) is followed by a quarter-wave plate (QWP). The latter is placed along the common path of the incident and scattered beams. 
At the incident path, it converts linearly polarized light to circularly polarized light. 
In the scattering path, an HWP and a linear polarizer are placed after the QWP. 
The HWP in the scattered path was rotated at different angles to obtain different helicities of the scattered light.  
Schematic diagrams for the experimental setups for linear and circular-polarization resolved Raman measurements are available in Figs.~\ref{fig.setup}(a) and~\ref{fig.setup}(b), respectively, in SM~\cite{Note1}.

\subsection{Computational details}
The first principles density functional theory (DFT) calculations were performed using the projected augmented-wave (PAW)~\cite{blochl1994} potentials implemented in the Vienna Ab initio Simulation Package (VASP)~\cite{kresse1993,kresse1996,kresse1999} code.
For the exchange-correlation energy, the generalized gradient approximation (GGA) in the Perdew--Burke--Ernzerhof (PBE) parametrization~\cite{perdew1996}. 
The energy cutoff for the plane-wave expansion was set to $400$~eV. 
The van der Waals (vdW) effect was included using the DFT-D3 method with the Becke--Johnson damping function~\cite{grimme2011}.
The correlation effects were introduced within the rotationally invariant DFT+U framework introduced by Liechtenstein {\it et al.}~\cite{liechtenstein1995}.
Due to the technical purpose (see Sec.~\ref{sec.system_para} in the SM~\cite{Note1}), the simplified unit cell with P3m1 symmetry was used in theoretical calculations.
The structure was optimized with $9 \times 9 \times 3$ ${\bm k}$-grid within the Monkhorst--Pack scheme~\cite{monkhorst1976}. 
As the convergence criterion for the optimization loop, we set the energy change to below $10^{-6}$ eV and $10^{-8}$ eV for ionic and electronic degrees of freedom, respectively.

In our DFT+U calculations, we assume an on-site Coulomb parameter $U=2.5$~eV and an exchange parameter $J=0.5$~eV of Fe $3d$ orbitals.
Obtained results (lattice parameters or Fe magnetic moments) theoretically well reproduce experimental parameters (see Sec.~\ref {sec.system_para} in the SM~\cite{Note1}).
However, we should note that in the literature, different values of $U$ and $J$ were assumed for FGT family members~\cite{zhu2016,joe.yang.19,ghosh2023}.

The dynamical properties were calculated using a $3 \times 3 \times 1$ supercell and a reduced $3 \times 3 \times 3$ ${\bm k}$-grid.
The phonon dispersion curves were obtained from 100 samples with random displacements of all atoms.
Then, the interatomic force constants (IFCs) were found using {\sc ALAMODE}~\cite{tadano2014}.
The irreducible representations were analyzed using the {\sc Phonopy} package~\cite{togo2023b,togo2023a}.
The Raman intensities were calculated for one octuple monolayer of F5GT using {\sc QERaman} software~\cite{hung.2024}, based on the {\sc Quantum ESPRESSO} code~\cite{giannozzi.20}.

\begin{figure}[!t]
\centering
\includegraphics[width=\linewidth]{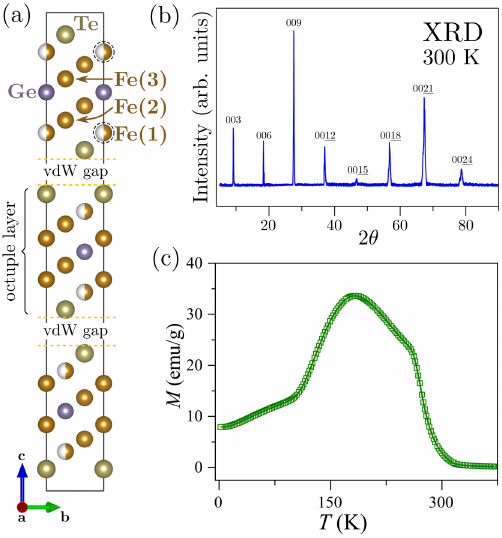}
\caption{
(a) Room temperature Fe$_{5}$GeTe$_{2}$ crystal structure,
realized rhombohedral R3m symmetry.
A conventional unit cell contains three octuple layers, separated by the van der Waals (vdW) gaps.
(b) Such a structure is confirmed by the X-ray diffraction (XRD) pattern obtained at $300$~K.
(c) shows the field-cooled (FC) magnetization vs. temperature plot of F5GT.
}
\label{fig.crystal}
\end{figure}

\section{Results and discussion}
\label{sec.results}

\subsection{Crystal structure and magnetic properties}

The FGT family members crystallize in a layered structure, which is typical for vdW materials.
F3GT realizes hexagonal P6$_{3}$/mmc symmetry (space group No.~194)~\cite{deiseroth2006}, while F4GT exhibits rhombohedral R$\bar{3}$m symmetry (space group No.~166)~\cite{wang.lu.23}.
 In both structures, a heterometallic slab of Fe and Ge is encapsulated by two Te layers, which are bonded to the adjacent layers through vdW force.

In the case of the F5GT crystal structure, it is more complicated, and there are several models describing this structure: rhombohedral R3m (space group No.~160)~\cite{stahl2018} and R$\bar{3}$m (space group No.~166)~\cite{may2019a} [see Fig.~\ref{fig.crystal}(a)].
In practice, F5GT is formed by incorporating an additional Fe layer into the F4GT structure.
However, due to the layer symmetry, a position close to the top or bottom of Te is indistinguishable.
This leads to three non-equivalent Fe positions, while Fe(1) has a refined fractional occupancy (around $0.5$)~\cite{may2019a}, and the system has R$\bar{3}$m symmetry [position marked by a black dashed circle on Fig.~\ref{fig.crystal}(a)].
Additional splitting arises from the adjustment of the Ge position to the occupied Fe(1) site.
As a consequence, the previously mentioned structural disorder emerges.
Summarized, the crystal structure is built from octuple layers formed by Te-\underline{Fe(1)}-Fe(2)-Fe(3)-Ge-Fe(3)-Fe(2)-Te atomic sandwiches.
Each octuple layer is shifted with respect to the previous one, resulting in ABC stacking.
In this case, three octuple layers are necessary to construct the conventional unit cell, while the split positions of Fe(1) and Ge atoms lead to structural disorder.

Such a crystal structure was examined by scanning transmission electron microscopy (STEM) and single crystal X-ray diffraction (XRD)~\cite{may2019a,zhang2020c,tian2020,zhang2024}.
In fact, the F5GT samples typically exhibit high quality; 
thus, XRD shows well-defined ($0 \; 0 \; 3l$) diffraction peaks characteristic of the rhombohedral structure [see Fig.~\ref{fig.crystal}(b)].

The unpolarized temperature-dependent field-cooled (FC) magnetization ($M$--$T$) plot, revealing  macroscopic magnetization of F5GT is presented in Fig.~\ref{fig.crystal}(c).
At a temperature around $310$~K, a ferromagnetic state emerges. Below $T_C$, the evolution of magnetization with temperature is complex. The anomaly around $274$~K can be linked to the formation of helical order~\cite{ly2021} or skyrmionic domains~\cite{schmitt2022,birch2024}; however, this remains under debate.
The broad cusp in the $110$--$260$~K temperature range reflects the competing energy scales of Dzyaloshinskii--Moriya interaction, Heisenberg exchange, magnetocrystalline anisotropy, and magnetostatic interactions~\cite{ly2021,fernandez2019,nembach2015}.
The drop in magnetization plot below $110$~K is induced by a change in stacking order along the $c$-axis~\cite{may2019a}.

\begin{figure*}[!t]
\centering
\includegraphics[width=\linewidth]{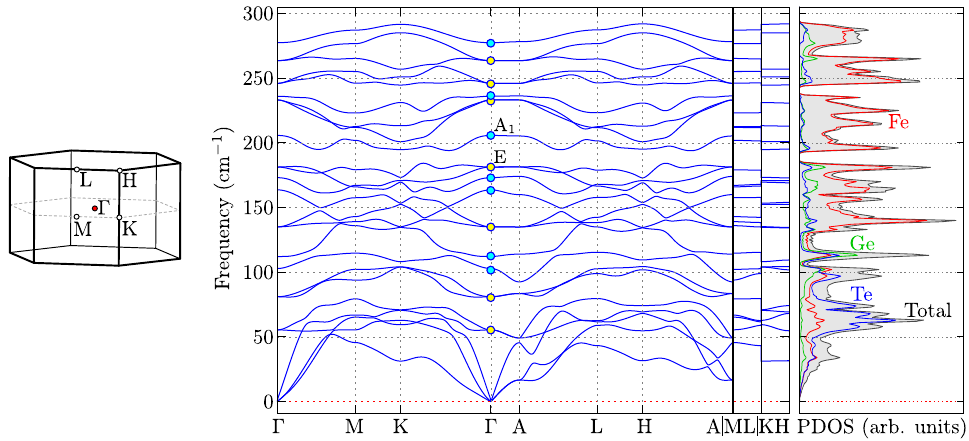}
\caption{
The phonon dispersion curves along high symmetry directions of the Brillouin zone (shown from left side) and phonon density of states.
Blue and yellow dots denoted A$_{1}$ and E modes at $\Gamma$ point.
Results obtained for F5GT model with P3m1 symmetry, within DFT+U calculations. 
}
\label{fig.ph_band}
\end{figure*}

\subsection{Lattice dynamics}

Due to the theoretical limitation of the system described with fractional occupancy, we investigate F5GT using a bulk model based on a single octuple layer periodically repeated along the $c$ direction.
Such a crystal with the trigonal P3m1 symmetry (space group No.~156) corresponds to the structure with AA stacking (see Sec.~\ref{sec.system_para} in the SM~\cite{Note1}).
Due to the weak vdW coupling between the octuple layers, we can assume that the obtained dynamical properties will be well followed by F5GT features.

The phonon dispersion and the phonon density of states for the proposed model are presented in Fig.~\ref{fig.ph_band}.
The obtained phonon spectrum does not exhibit imaginary soft modes, and the system is stable in a dynamical sense.
Acoustic branches exhibit a non-linear dependence around the $\Gamma$ point, typical for layered systems~\cite{mobaraki2019,wilczynski2022,bae2022}. 
Such behavior is mostly associated with transverse out-of-plane acoustic vibrations~\cite{nika2009,libbi2021} (see Fig.~\ref{fig.irr_gamma} in the SM~\cite{Note1}).
The phonon spectrum is dominated by the vibrations of Fe atoms.
The Te atom vibrations are mostly limited to the lower frequency range (below $125$~cm$^{-1}$), while Fe atom vibrations are primarily expected in the higher frequency range (above $125$~cm$^{-1}$).
Modes associated with Ge atom vibrations are realized within intermediate frequencies (around $125$~cm$^{-1}$), and are well mixed with other modes.

\subsection{Raman spectroscopy and electron--phonon coupling}

\begin{figure}[!t]
\centering
\includegraphics[width=\linewidth]{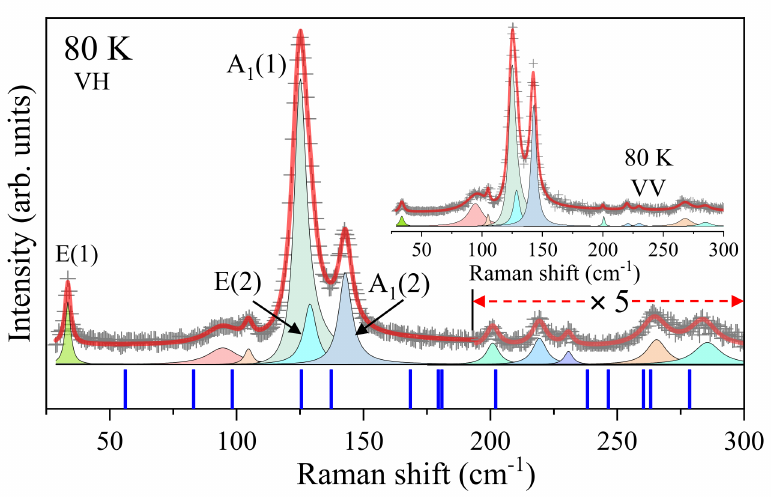}
\caption{
Raman spectrum measured at $80$ K in VH (inset: VV) polarization configuration. While the ``+'' symbols are measured data points, the net fitted curves using Eq.~(\ref{eq.fit}) are shown by red curves. 
Deconvoluted components are presented as color-shaded areas.
Dominant modes in VH configuration ---A$_{1}$(1), A$_{1}$(2), E(1), and E(2)--- are labeled.
Theoretically obtained modes are marked by blue bars below. 
\label{fig.raman}} 
\end{figure}

{\it Raman spectrum}---
For the proposed model with P3m1 symmetry, the optical phonon modes at the $\Gamma$ point can be decomposed into irreducible representations as $7 \text{A}_{1} + 7 \text{E}$.
All optical modes are simultaneously IR- and Raman-active.
Modes A$_1$ (E) describe out-of-plane (in-plane) vibrations (see Fig.~\ref{fig.irr_gamma} in the SM~\cite{Note1}).
The Raman spectra obtained for $80$~K for VH and VV polarization configurations are shown in Fig.~\ref{fig.raman} (see also Sec.~\ref{cleave} in SM~\cite{Note1}).
Of the 14 expected active modes, we observed 11 over the spectral range from $30$ to $300$~cm$^{-1}$.
The spectral region between $110$ and $160$~cm$^{-1}$, with strong Raman modes, could not be fitted using only Lorentzian functions for all modes (see Fig.~\ref{fig.Fano_fit} in the SM~\cite{Note1}).
Thus, to fit the entire spectral range in Fig.~\ref{fig.raman}, we used
\begin{eqnarray}
\nonumber I ( \omega ) = \sum_{i} \frac{A_i}{2\pi} \frac{ \Gamma_{i} }{ (\omega - \omega_{i})^{2} + (\Gamma_{i}/2)^{2} } + \sum_{j}\frac{F_{j} (q_{j}+\epsilon_{j})^2 }{ (1+\epsilon_{j}^2) } . \\
\label{eq.fit} 
\end{eqnarray}
The first term corresponds to nine Raman modes, with Raman shift $\omega_{i}$, width $\Gamma_{i}$, and integral intensity $A_{i}$. 
Two other Raman modes around $125$~cm$^{-1}$ and $143$~cm$^{-1}$ could be best fitted with Fano line shapes (see Sec. \ref{Fano_fit}), given by the second term, where $F_j$ is a constant, while $\epsilon_{j} = ( \omega_{j} - \omega_{0}^{j} ) / \Gamma_j$, where $\omega_{0}^{j}$ and $\Gamma_j$ are the peak positions and widths, respectively, and $q_j$ is the asymmetry parameter.
The best fit to the data points is represented by the red solid curve, and the deconvoluted components are indicated by the color-shaded peaks in Fig.~\ref{fig.raman}.

The most dominant Raman active modes at frequencies $124.8$, $142.9$, $33.6$, and $128.9$~cm$^{1}$ are marked as A$_1$(1), A$_1$(2), E(1), and E(2), respectively.
Characteristic frequencies are close to those observed for another FGT compound~\cite{milosavljevic2019,du2019,xu2020b,
kong2021,dang2023,zhang2025b}, and are in agreement with the frequencies of modes obtained for (un)doped F5GT~\cite{yadav2024,li2025,yu2025,nair.mallik.25}.
The theoretically obtained frequencies of the modes at $\Gamma$ point are available in (see Tab.~\ref{tab.irr_gamma} in the SM~\cite{Note1}).
The difference in Raman spectra can be related to the structural disorder within Fe sublattice~\cite{yadav2024,li2025}. 
This hypothesis can be supported by the theoretically obtained phonon DOS, where vibrations of Fe atoms are realized within whole range of frequencies (Fig.~\ref{fig.ph_band}, and Fig.~\ref{fig.irr_gamma} in the SM~\cite{Note1}).
Thus, any modification to the Fe sublattice directly affects phonon modes.

\begin{figure*}[!t]
\centering
\includegraphics[width=\linewidth]{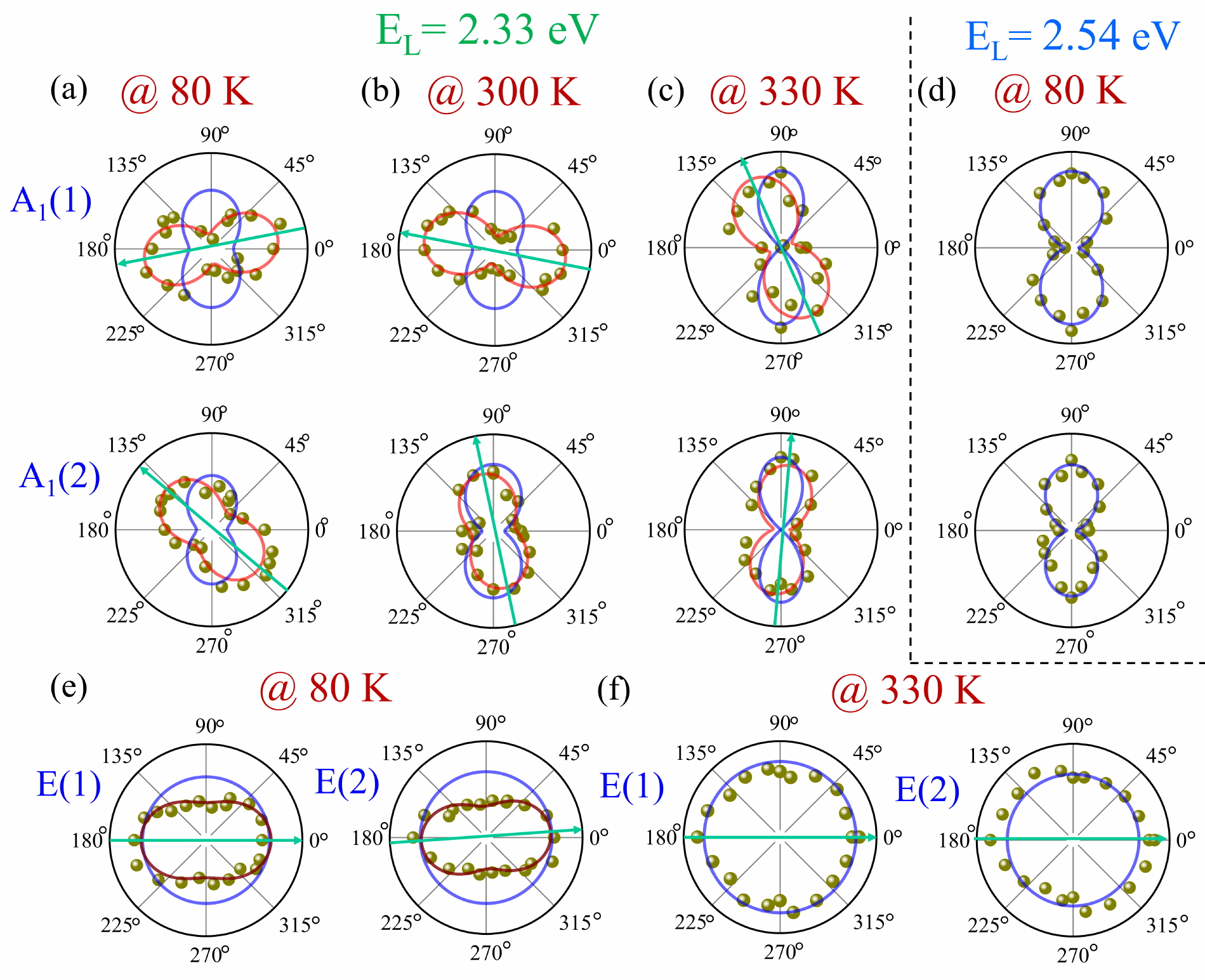}
\caption{
Polar plots of normalized Raman intensity for the discussed modes (as labeled) with excitation energy $2.33$~eV at $80$~K in (a) at $300$~K in (b) and at $330$~K in (c), same with excitation energy $2.54$~eV at $80$~K in (d), normalized intensity for both E modes are at $80$~K in (e) at $330$~K in (f) respectively,
for the vertical polarization of the incident light on the sample. 
In (a)-(f), the blue curves correspond to the angle-resolved intensity for A$_{1}$ mode and E mode [Eq.~(\ref{eq.da1}) and~(\ref{eq.dephi}) in the SM~\cite{Note1}, respectively] using the Raman tensor in Eq.~(\ref{eq.raman_tensor_mag}). 
Red curves are the best fit to the data points using Eq.~(\ref{eq.da1phi}) in the SM~\cite{Note1} for A$_{1}$ modes following the tensor in Eq.~(\ref{eqn_DAmod}).
In (e), maroon curves correspond to best fit to the data points using using Eq.~(\ref{eq.dephi2}) in the SM~\cite{Note1}.
\label{fig.angle_raman}}
\end{figure*}

{\it Angle-dependent Raman intensity.}---
Linear polarization-dependent Raman spectra recorded at $80$~K are available in Fig.~\ref{fig.full_angle_polar} in the SM~\cite{Note1}.
According to the Placzek approximation, the Raman intensity is given by $I \propto \left| \hat{e}_{s}^{T} \cdot R \cdot \hat{e}_{i} \right|^2$, where $\hat{e}_i$ and $\hat{e}_s$ are the polarization unit vectors of the incident and scattered light, respectively.
For the modeled crystal structure, the Raman tensors in the crystal coordinate system are:
\begin{eqnarray}
\label{eq.raman_tensor}
R(\text{A}_{1}) &=& \left( \begin{array}{ccc}
    a & 0 & 0 \\
     0 & a & 0 \\
     0 & 0 & b \\
    \end{array}\right), \\
\nonumber R(\text{E}^\text{I}) &=& \left( \begin{array}{ccc}
    0 & c & d \\
     c & 0 & 0 \\
     d & 0 & 0 \\
    \end{array}\right), \quad \text{and} \quad
R(\text{E}^\text{II}) = \left( \begin{array}{ccc}
    c & 0 & 0 \\
     0 & -c & d \\
     0 & d & 0 \\
     \end{array}\right).
\end{eqnarray}
Due to the selection rules, the A$_1$ and E modes can be easily recognized by angle-resolved linear or circular polarization-dependent Raman measurements (see Sec.~\ref{sec.raman_intens}(b) in the SM~\cite{Note1}).
In the laboratory coordinate system, under linearly polarized incident radiation the A$_1$ modes exhibit strong 
polarization-dependence, while E modes are polarization-independent (see Eqn.~\ref{normal_tensor}
Sec.~\ref{sec.Raman_tensors} in the SM~\cite{Note1}).
However, the polar plots of intensity variation in our angle-resolved polarization-dependent Raman measurements deviate markedly from the expected behavior (Fig.~\ref{fig.angle_raman}). 
For modes A$_{1}$(1) and A$_1$(2), we observe a tilt in the polar plots. This deflection is around $\varphi \simeq 10^{\circ}$ and $\sim 137^{\circ}$, respectively.
Such an effect can be explained in several ways.

First, we look into the possible role of magnetism. 
The tilt or distortion of the polar plot of the Raman intensity pattern can be influenced by an external magnetic field, as observed in several compounds~\cite{liu2020,mccreary2020,wan2021,sun2025}.
In the F5GT, with the internal magnetism, a similar tilt in the polar plots of Raman intensity can be expected.
In the magnetic crystal, new non-zero complex elements should be added to the Raman tensors. For the A$_1$ modes it can be written as ~\cite{cracknell1969,lyu2020}:
\begin{eqnarray}
\label{eq.raman_tensor_mag}
\text{DA}_{1} &=& \left( \begin{array}{ccc}
    a & if & 0 \\
     -if & a & 0 \\
     0 & 0 & b \\
    \end{array}\right)
\end{eqnarray}
(detailed discussion can be found in Sec.~\ref{sec.Raman_tensors} in the SM~\cite{Note1}).
Considering the modified Raman tensors, the expected angular variations of intensity of the
modes are shown by blue curves, which do not explain the observed results for the A$_{1}$ and E modes.

To explain the observed intensity variation, additional modifications to the Raman tensor can be introduced due to electron--phonon coupling.
From the microscopic viewpoint of quantum mechanical theory of Raman scattering~\cite{jiang.saito.07,han2022,lin2019,pimenta2021}, using the third-order time-dependent perturbation theory, the Raman tensor elements can be written as~\cite{pimenta2021}:
\small
\begin{eqnarray}
R_{ij}^{\mu \nu} \propto 
\sum_{g,n,n'} \sum_{\vec{k}} 
\frac{
\langle \psi_g(\vec{k}) | \vec{E}^j \!\cdot\! \vec{d} | \psi_{n'}(\vec{k}) \rangle 
\langle \psi_{n'}(\vec{k}) | H_{\mathrm{el-ph}}^{\mu} | \psi_n(\vec{k}) \rangle
}
{
\left[E_L - E_{ng}(\vec{k}) + i\Gamma_n \right]
}
\nonumber \\
\times
\frac{
\langle \psi_n(\vec{k}) | \vec{E}^i \!\cdot\! \vec{d} | \psi_g(\vec{k}) \rangle
}
{
\left[E_L - E_{\mathrm{ph}}^{\mu} - E_{n'g}(\mathbf{k}) + i\Gamma_{n'} \right]
}. \nonumber \\
\label{Raman_tensor}
\end{eqnarray}
The sums in Eq.~(\ref{Raman_tensor}) are performed over all electronic wavevectors \textbf{k} within the first Brillouin zone and over the electronic branches in ground state $g$ and intermediate states $n$ and $n^\prime$. 
At first and third interaction vortex, for the electron-radiation Hamiltonian $H_{el-rad} = -\vec{d}\cdot\Vec{E}$, $\vec{d}$ is the electric-dipole operator and $\vec{E}$ is the electric field of the radiation. 
$H_{el-ph}$ is the electron--phonon interaction Hamiltonian. 
In the denominator  $E_{n^{\prime}g}(\vec{k}) = E_{n^{\prime}}(\vec{k}) - E_g(\vec{k})$, and $E_{ng}(\vec{k}) = E_n(\vec{k}) - E_g(\vec{k})$, are the energy differences between the ground and intermediate states at a given wavevector \textbf{k}, with the damping constants $\Gamma_n$ and $\Gamma_{n^{\prime}}$ corresponding to the finite lifetime of photo excited states. Under resonance, the complex value of the electronic susceptibility $\chi(\omega)=\chi^{\prime}(\omega)+\chi^{\prime \prime}(\omega)$, gives rise to complex electron--photon and electron--phonon matrix elements, and hence, the fully complex Raman tensor.
As the Raman tensor depends on the excitation energy $E_L$, in the case of strong coupling for a specific resonant excitation energy, an additional complex phase can be introduced on the diagonal elements of the Raman tensor~\cite{pimenta2021}.
Thus, we modify the complex Raman tensor for A$_{1}$ mode in the form with minimal change in the same in Eq.~(\ref{eq.raman_tensor_mag}):
\begin{eqnarray}
\text{DA}_1^\text{mod} = \left( \begin{array}{ccc}
	a e^{i\alpha_{a}} & if & 0 \\
    -if & a e^{i\alpha_{a}} & 0 \\
    0 & 0 & b e^{i\alpha_{b}} 
    \end{array}\right) .
    \label{eqn_DAmod}
\end{eqnarray}
A similar complex phase can also be added to the Raman tensor to describe the E mode.
The intensity obtained for A$_{1}$ modes using DA$_1^\text{mod}$ is presented by red contours in Fig.~\ref{fig.angle_raman}(a), for the angle resolved intensity variation at 80 K, and it relatively well describes our results (fitted parameters are available in Tab.~\ref{tab.weight_factors} in
the SM~\cite{Note1}).
Theoretically obtained angle-dependent complex Raman intensities for such Raman tensors in the cases of one-octuple F5GT layer are presented in Fig.~\ref{fig.angle_raman_theo} in the SM~\cite{Note1}.
The tilt of the maximal intensity corresponds fairly well to what was observed experimentally (cf.~Fig.~\ref{fig.angle_raman}), and is also reported for layered dichalcogenides~\cite{saito2016,kumar2020,pimenta2021,resende2021}.
The modification of the obtained Raman intensities from the expected ones shows a strong anisotropy of the electron--phonon coupling.

\begin{figure}[!t]
\centering
\includegraphics[width=0.8\linewidth]{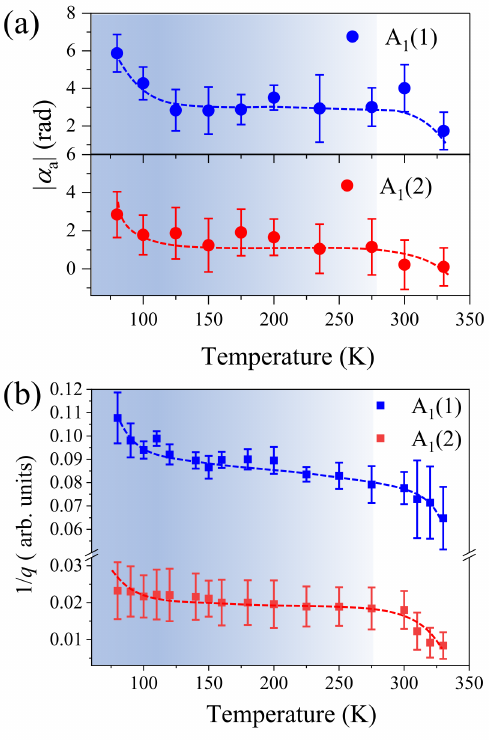}
\caption{Temperature variation of the (a) magnitude of the phase factor $\alpha_a$ introduced in Raman tensor, reflecting the non-trivial nature of the scattering amplitude arising from electron--phonon coupling, and (b) electron--phonon coupling strength, $1/q$ for the A$_1$ modes in VV polarization configuration. The blue shaded area marks the temperature range for the magnetically ordered phase.
\label{fig.fano_q-1}}
\end{figure}

Angle-resolved Raman intensity has been recorded at various temperatures and analyzed using Eq.~(\ref{eqn_DAmod}) for all plots over the entire temperature range. Fig.~\ref{fig.angle_raman}(b) and \ref{fig.angle_raman}(c) plot the angle-resolved Raman intensity recorded at two characteristic higher temperatures, $300$~K and $330$~K, respectively. It is interesting to note that the tilt in the plots from the expected one (blue curve) persists even at 330 K. As mentioned earlier, such a tilt can be only explained with the off-diagonal terms in the Raman tensor and the addition of a phase factor to the Raman tensor element. To minimally modify the analysis protocol, we used a similar tensor described in Eq.~(\ref{eqn_DAmod}), except $f$ instead of $if$ as off-diagonal elements,
to fit the data points of angle-resolved intensity variation beyond $T>T_C$. However, in the paramagnetic phase, the off-diagonal terms in the Raman tensor for the plots recorded for $T>T_C$, reflect a dynamical mixing of A$_1$
 and E symmetry channels, likely mediated by spin--orbit coupled spin fluctuations persisting above $T_C$~\cite{Bera2024}.
Below $T_C$, spin--orbit interaction~\cite{kong2024} links the ordered magnetization to the lattice, lowering the symmetry to a magnetic point group and allowing antisymmetric, magnetization-dependent components in the Raman tensor. In other words, below $T_C$, spin--orbit coupling locks static magnetization direction $\vec{M}$ to the lattice~\cite{Bera2024,Liu2023}.
Above $T_C$, although the time-averaged magnetization vanishes, however, spin--orbit coupling converts short-range spin correlations into an effective modification of the Raman susceptibility, resulting in symmetry mixing \cite{Liu2023,kong2024} and apparent off-diagonal tensor elements even for the mode A$_1$.  The temperature variation of the magnitude of the phase factor, $|{\alpha_a}|$, is shown in Fig.~\ref{fig.fano_q-1}(a). The value of $\alpha$ drops with the rise in temperature. The above discussed arguments suggest that below $T<T_C$, the decrease in the value of $\alpha_a$ is governed by the drop in electron--phonon coupling with the rise in temperature in the magnetic ordered phase, whereas, for $T>T_C$ it reflects the same, in the presence of spin--orbit coupling under fluctuating short-range magnetic interaction.

To further establish the tilt in polar plots of Raman intensity of the A$_1$ modes in Fig.~\ref{fig.angle_raman} to electron--phonon coupling under optical resonance, we carried out wavelength-dependent Raman measurements. 
Raman spectra were recorded for excitation energies of the laser source at $2.18$, $2.33$, $2.41$, and $2.54$~eV. For 2.54 eV as the excitation energy, the drop in electron--phonon coupling strength, as obtained from the value of $1/q$ from spectral analysis  (see Fig.~\ref{fig.fano_asymmetry_energy} in the SM~\cite{Note1}), prompted us to choose this probing energy to study the tilt in the polar plot of the intensity of the A$_1$ mode in off-resonance condition.
Indeed, we obtained symmetric polar plots of the same without any tilt for this excitation energy (see Figs.~\ref{fig.angle_raman} (d). Here we would like to mention that the measured value of the Fano parameter $q$ of the A$_1$(1) mode is similar to that reported for other vdW magnets such as. Cr$_2$Ge$_2$Te$_6$~\cite{Chakkar2024} and Cr$_{1.45}$Te$_2$~\cite{Lan2025}.

We note that the microscopic origin of the observed resonance in the Raman intensity is not fully understood. Here, we can mention several mechanisms:
{\it (i)} interband transition (e.g., Fe $3d$ to higher bands) energies in metallic F5GT lie near photon energies~\cite{ZHANG2025c,ershadrad2022}. 
The overlap of laser excitation energy with such transition energy can give rise to a resonance phenomenon.
{\it (ii)} The local plasmonic effect in metallic F5GT can amplify the coupling locally~\cite{brennan2024}. {\it (iii)} spin--orbit coupling and magnetism~\cite{Bera2024} can modify optical matrix elements, resulting in an enhanced effect of the electron--phonon interaction.

Fig.~\ref{fig.angle_raman}(e) plots the angle-resolved intensity of the E1 mode at 80 K using resonance $E_L$=2.33 eV. For the incoherent superposition of two degenerate E modes, its angle-resolved Raman intensity is expected to be isotropic (as shown by blue curves).
Several articles in the literature discuss the Raman generation of coherent phonons in the context of ultrafast phonon dynamics~\cite{ishioka2013,stevens2002}, especially under resonance conditions; however, such a physical concept of the coherent superposition of tensors is rare in the context of spontaneous Raman scattering. The observed intensity of the polar plot of the E modes could be explained [see note of Eq.~(\ref{eq.dephi2}) in the SM~\cite{Note1}] by assuming such a coherent superposition of two degenerate E modes in the present case, which vanishes at high temperature at $330$~K [see Fig.~\ref{fig.angle_raman}(f)].

\begin{figure*}[!t]
\centering
\includegraphics[width=\linewidth]{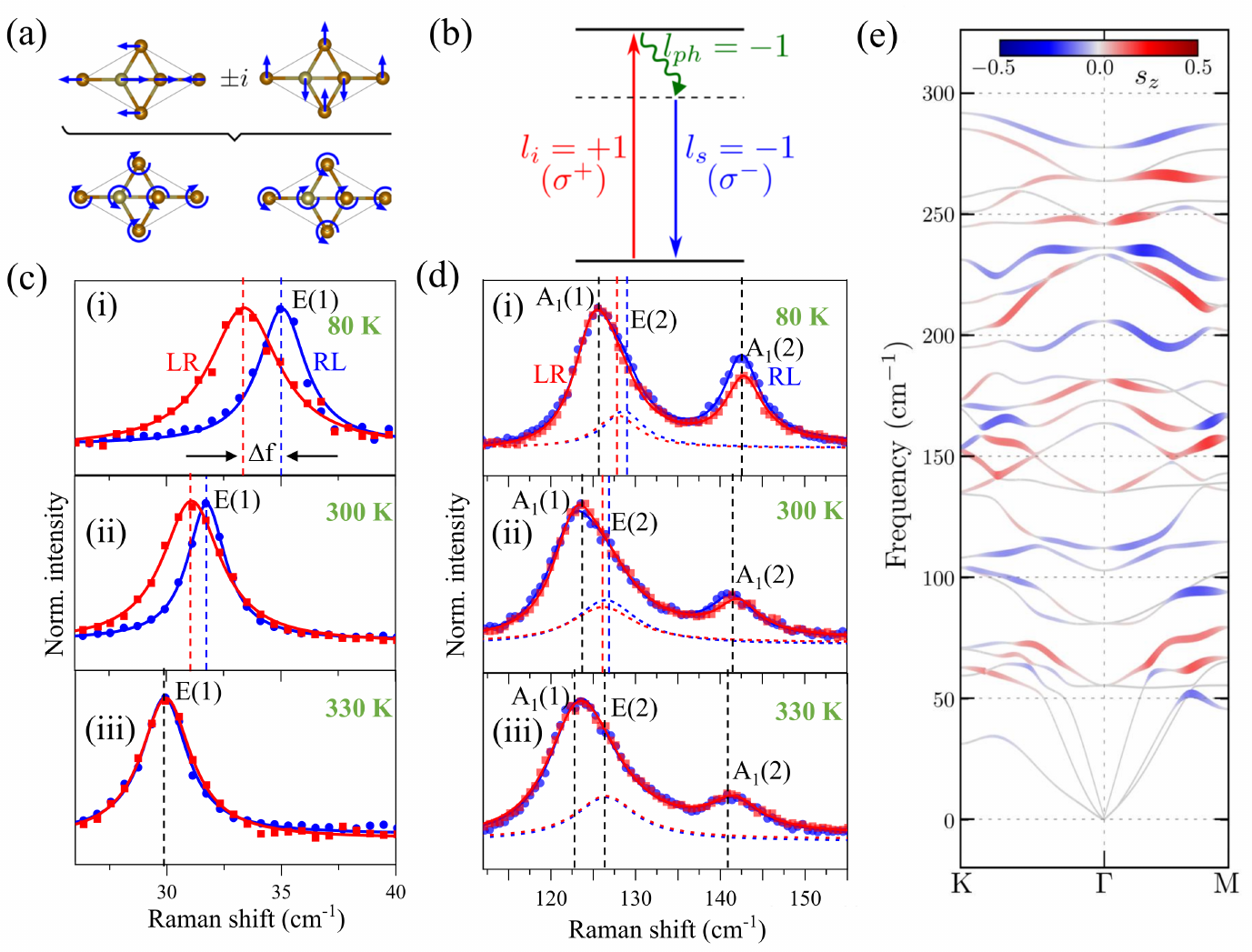}
\caption{
(a) Superposition of double-degenerated modes E results in right-handed and left-handed circular motions at the $\Gamma$ point.
(b) Schematic of the helicity of the photon switched by the chiral phonon.
(c,d) The helicity-resolved Raman spectra of F5GT at $80$~K, $300$~K, and $330$~K (as labeled).
RL and LR configurations are presented by blue and red lines/points, respectively, 
(e) Theoretically calculated the phonon circular polarization.
Colored fat-bands correspond to the branches realizing the chiral phonons. 
}
\label{fig.chiral}
\end{figure*}

{\it Fano parameter.}---
The possible electron--phonon coupling can also be investigated via the Fano parameters $q_j$ in Eq.~(\ref{eq.fit}), which reflects the interference of the discrete phonon with the continuum of electronic states.
For the Raman peaks A$_{1}$(1) and A$_{1}$(2), we found the Fano parameters $q \sim 8$--$13$ and $35$--$50$, respectively ($q$ values).
Positive values indicate constructive (in-phase) coupling of this mode with the electron continuum.
The electron--phonon coupling strength is given by the dimensionless parameter $1/q$, and decreases with increasing temperature [see Fig.~\ref{fig.fano_q-1}(b), plotted for VV polarization configuration]. As both $|\alpha_a|$ and $1/q$ relate to the integral intensity of the Raman mode, it is not surprising to observe that both follow a similar trend with temperature.
The monotonic drop in $1/q$ and $\alpha_a$ with temperature, instead of following the modulation of $M(T)$ in Fig.~\ref{fig.crystal}(c), establishes a weak connection between electron--phonon coupling and magnetic order.

Together, these observations demonstrate that optical resonance-induced electron--phonon coupling in F5GT is both anisotropic and magnetically sensitive, providing a microscopic basis for the interplay among lattice, electronic, and magnetic degrees of freedom in this layered van der Waals system.

\subsection{Chiral phonons}

The three-folded rotation around $z$ axis ($C_{3z}$) is one from the group generators of the proposed F5GT model with P3m1 symmetry, as well as of rhombohedral models with R3m or R$\bar{3}$m symmetries.
Thus, at high symmetry points, which are invariant under $C_{3}$ rotation, we can obtain $C_{3} u_{\bm k} = \exp \left( - i (2\pi/3) l_\text{ph} \right) u_{\bm k}$, where $l_\text{ph} \in \{-1 , 0 , 1 \}$ is the pseudoangular momentum (PAM) of a phonon with wave function $u_{\bm k}$~\cite{zhang2015}.
States with non-zero $l_\text{ph} = \pm 1$ correspond to the chiral phonons describing the circular motion of atoms around the equilibrium position. 
There are several ways to investigate chiral phonons, for example: 
(i) the experimental study of cross-circular Raman scattering (RL and LR), and (ii) the theoretical study of phonon circular polarization.

At the Brillouin zone center, the total PAM, by definition, is zero~\cite{zhang2015}.
However, the E modes can be understood as a superposition of the right-handed and left-handed polarized modes, with opposite chirality [Fig.~\ref{fig.chiral}(a)].
Such modes can be decoupled in the presence of an external magnetic field and recorded as separate peaks within Raman spectroscopy experiments~\cite{schaack1976}.
In our case, the internal magnetic order plays the role of the effective magnetic field as observed for the Weyl semimetall
 ic compound Co$_3$Sn$_2$S$_2$ ~\cite{che2025}.
The conservation of PAM in the Raman process dictates that $l_s = l_i - l_{ph}$, where $l_i$, $l_s$, and $l_{ph}$ are the angular momenta of the incident light, scattered light, and the phonon involved in the Raman process. 
For example, the incident right-handed polarized photon emits a left-handed phonon and then scatters into a left-handed polarized photon [Fig.~\ref{fig.chiral}(b)].
Since Raman scattering involving E modes switches the polarization of circularly polarized light, this mode must have a non-zero PAM.
Simultaneously, a similar process is forbidden in the case of A$_{1}$ mode.

Indeed, we observed the splitting of the E modes within cross-circular (RL and LR) helicity resolved Raman scattering.
Such splitting ($f$) is pronounced for the E(1) mode at 80 K.
The same for the observed peaks at $f_\text{RL} = 35.0$~cm$^{-1}$ and $f_\text{LR} = 33.4$~cm$^{-1}$ for RL and LR configurations, respectively, corresponds to the Raman shift difference $\Delta f = f_\text{RL}-f_\text{LR} = 1.6 $~cm$^{-1}$ at 80 K [see Fig.~\ref{fig.chiral}(c)(i)].
For E(1), such splitting $\Delta f$ decrease with the increase in temperature, and vanishes at $330$~K in the paramagnetic phase of F5GT (cf. Fig.~\ref{fig.chiral}(c) (ii) and (iii), and see Fig.~\ref{E mode_split_temp} in the SM~\cite{Note1}).
This directly shows the interplay between the chiral phonon and the F5GT intrinsic magnetic order.
However, for E(2) it is nontrivial to confirm the observed splitting of $0.7$~cm$^{-1}$, as it lies at the shoulder of the strong A$_1$(1) mode [see Fig.~\ref{fig.chiral}(d)].
This clearly shows two types of phonon modes: one is only Raman active in the RL configuration, while the second is only active in the LR configuration.
Similar results were previously reported in several compounds, such as $\alpha$-quartz~\cite{oishi2024}, Te~\cite{zhang2023,ishito2023,spirito2024}, and $\alpha$-HgS~\cite{ishito2023b}, which, due to their chiral crystal structure, naturally realize chiral phonons.
More recently, such behavior was reported in the ferromagnetic Co$_{3}$Sn$_{2}$S$_{2}$ without a chiral structure~\cite{che2025}. 
It is also noteworthy that the out-of-plane non-degenerate A$_1$ modes in Fig.~\ref{fig.chiral}(d) do not split in the said circular polarization configurations, as they do not carry PAM.

This observation can be supported by the phonon circular polarization obtained theoretically.
To study the circular motion in, e.g., $xy$ plane, we can define a new basis of the phonon polarization vectors:
$R_{i} = \sfrac{1}{\sqrt{2}} \left(\cdots 1 \; i \; 0 \; \cdots \right)^{T}$, $L_{i} = \sfrac{1}{\sqrt{2}} \left(\cdots 1 \; -i \; 0 \; \cdots \right)^{T}$, $Z_{i} = \left(\cdots 0 \; 0 \; 1 \; \cdots \right)^{T}$, where $R_{i}$ and $L_{i}$ vectors describe right-handed and left-handed circular polarization for the $i$th atom ($Z_{i}$ as the vector perpendicular to the circulation plane is unchanged). 
Then, every polarization vector $\varepsilon$ can be represented as $\varepsilon = \sum_{i} \varepsilon^{R}_{i} R_{i} + \varepsilon^{L}_{i} L_{i} + \varepsilon^{z}_{i} Z_{i}$, where $\varepsilon^{o}_{i} = \langle o_{i} \vert \varepsilon \rangle$.
Thus, the phonon circular polarization is given as $s_{z} = \hbar \sum_{i} \vert \varepsilon^{R}_{i} \vert^{2} - \vert \varepsilon^{L}_{i} \vert^{2}$.
Thus, $s_{z} \neq 0$ for all phonon branches with non-compensated right- and left-handed circulations.
The results of our calculations are presented in Fig.~\ref{fig.chiral}(e) (see also Fig.~\ref{fig.full_circ} in the SM~\cite{Note1}).
As we can see, along some directions, the non-vanishing total $s_{z}$ exists. 
The extreme values of $s_{z}$ correspond to the ideal circular motion of atoms, while $0 < \vert s_{z} \vert < 0.5$ correspond to the motion along ellipses.
In fact, the chiral phonons in the F5GT realize only elliptic orbits.
Moreover, such orbits are not realized exactly within $xy$ planes.
This feature is directly associated with the crystal structure of F5GT.
In the case of the symmetric structure, the circulation can be realized directly within the $xy$ plane.
This was reported earlier in the highly symmetric (Mn/Fe)Bi$_{2}$Te$_{4}$ magnetic topological insulators with R$\bar{3}$m symmetry~\cite{kobialka2022}.
Those systems possess inversion symmetry, and there are always pairs of similar atomic layers.
In the case of the F5GT system, the symmetry is broken due to the asymmetry of the Fe atomic layer with respect to the central Ge layer.
The absence of inversion symmetry leads to non-vanishing total phonon circular polarization~\cite{coh2023}.
The chiral phonons can be realized close to the $\Gamma$ point across the entire range of frequencies.
Thus, it cannot be associated with any specific atomic sublattice of F5GT.
Visualization of the chiral phonons is presented in Fig.~\ref{fig.trace} in the SM~\cite{Note1}.

\section{Conclusions}
\label{sec.sum}

In summary, we discuss the anisotropic electron--phonon coupling and chiral phonons in the vdW F5GT system.
Our study was based on experimental Raman spectroscopy and theoretical techniques based on {\it ab initio} calculations.
The existence of strong anisotropic electron--phonon coupling is reflected in the modification and tilt of the angular-dependent Raman intensity measured experimentally, and is further supported by direct calculations of the same from the Raman tensors.
Due to the threefold rotation axis, in F5GT, the chiral phonons associated with the circular motion of the atoms around their equilibrium positions can exist.
Identification of such exotic phonons was possible from the peak splitting of the doubly degenerate E modes under the cross-circular RL and LR Raman measurements.
This finding is also supported by theoretically calculated atomic trajectories under circularly polarized light, and also from the phonon dispersion, which reveals branches with non-vanishing total pseudoangular momentum.

The above findings establish F5GT not only as an interesting material from the perspective of possible applications in electronic nanodevices, but also as a compelling platform for exploring phonon-based functionalities.
The realization of chiral phonons opens the way to possible phonon-based engineering, e.g., exploring the possible phonon thermal Hall effect~\cite{zhang2015}.
Our results identify F5GT as an ideal system for investigating emergent couplings among lattice, electronic, and magnetic degrees of freedom and for advancing the understanding of chiral phonons in magnetic van der Waals materials.

\begin{acknowledgments}
Some figures in this work were rendered using {\sc Vesta}~\cite{momma.izumi.11} software.
We gratefully acknowledge Polish high-performance computing infrastructure PLGrid (HPC Centers: ACK Cyfronet AGH) for providing computer facilities and support within computational grant no. PLG/2026/019231
\end{acknowledgments}

\bibliography{ref}

@article{chen2023,
  title={Thermal cycling induced alteration of the stacking order and spin-flip in the room temperature van der {Waals} magnet {Fe$_{5}$GeTe$_2$}},
  author = {Chen, Xiang and Tian, Wei and He, Yu and Zhang, Hongrui and Werner, Tyler L. and Lapidus, Saul and Ruff, Jacob P. C. and Ramesh, Ramamoorthy and Birgeneau, Robert J.},
  journal={Phys. Rev. Mater.},
  volume={7},
  number={4},
  pages={044411},
  year={2023},
  publisher={APS},
 doi = {10.1103/PhysRevMaterials.7.044411},
  url = {https://doi.org/10.1103/PhysRevMaterials.7.044411}
}

@article{yadav2024,
  title={Thermal history-dependent characteristics in van der {Waals} ferromagnet {Fe$_{5-x}$GeTe$_2$} ($x \sim 0.16$)},
  author={Yadav, Ramesh Lalmani and Bag, Pallab and Lai, Chien-Chih and Kuo, Yung-Kang and Kuo, Chia-Nung and Lue, Chin-Shan},
  journal={APL Mater.},
  volume={12},
  number={8},
  year={2024},
pages = {081103}, 
  publisher={AIP Publishing},
doi = {10.1063/5.0215121},
url = {https://doi.org/10.1063/5.0215121}
}

@article{yang2021,
  title = {Strong magneto-optical effect and anomalous transport in the two-dimensional van der {Waals} magnets {Fe$_{n}$GeTe$_{2}$} ($n=3$, 4, 5)},
  author = {Yang, Xiuxian and Zhou, Xiaodong and Feng, Wanxiang and Yao, Yugui},
  journal = {Phys. Rev. B},
  volume = {104},
  issue = {10},
  pages = {104427},
  numpages = {15},
  year = {2021},
  month = {Sep},
  publisher = {American Physical Society},
  doi = {10.1103/PhysRevB.104.104427},
  url = {https://doi.org/10.1103/PhysRevB.104.104427}
}

@article{lyu2020,
  title={Probing the ferromagnetism and spin wave gap in {VI$_3$} by helicity-resolved {Raman} spectroscopy},
  author={Lyu, BingBing
and Gao, YiFan
and Zhang, Yujun
and Wang, Le
and Wu, Xiaohua
and Chen, Yani
and Zhang, Jiasheng
and Li, Gaomin
and Huang, Qiaoling
and Zhang, Naipeng
and Chen, Yuanzhen
and Mei, Jiawei
and Yan, Hugen
and Zhao, Yue
and Huang, Li
and Huang, Mingyuan},
  journal={Nano Lett.},
  volume={20},
  number={8},
  pages={6024},
  year={2020},
  publisher={ACS Publications},
DOI = {10.1021/acs.nanolett.0c02029},
URL = {https://doi.org/10.1021/acs.nanolett.0c02029}

}

@article{kobialka2022,
  title={Dynamical properties of the magnetic topological insulator {$T$Bi$_2$Te$_4$} ({$T=$ Mn, Fe}): Phonons dispersion, {Raman} active modes, and chiral phonons study},
  author={Kobia{\l}ka, Aksel and Sternik, Ma{\l}gorzata and Ptok, Andrzej},
  journal={Phys. Rev. B},
  volume={105},
  number={21},
  pages={214304},
  year={2022},
  publisher={APS},
doi = {10.1103/PhysRevB.105.214304},
  url = {https://doi.org/10.1103/PhysRevB.105.214304}
}

@article{du2019,
  title={Lattice dynamics, phonon chirality, and spin--phonon coupling in {2D} itinerant ferromagnet {Fe$_3$GeTe$_2$}},
  author = {Du, Luojun and Tang, Jian and Zhao, Yanchong and Li, Xiaomei and Yang, Rong and Hu, Xuerong and Bai, Xueyin and Wang, Xiao and Watanabe, Kenji and Taniguchi, Takashi and Shi, Dongxia and Yu, Guoqiang and Bai, Xuedong and Hasan, Tawfique and Zhang, Guangyu and Sun, Zhipei},
  journal={Adv. Funct. Mater.},
  volume={29},
  number={48},
  pages={1904734},
  year={2019},
  publisher={Wiley Online Library},
  DOI={10.1002/adfm.201904734},
  URL={https://doi.org/10.1002/adfm.201904734}
}

@article{zhang2015,
  title = {Chiral Phonons at High-Symmetry Points in Monolayer Hexagonal Lattices},
  author = {Zhang, Lifa and Niu, Qian},
  journal = {Phys. Rev. Lett.},
  volume = {115},
  issue = {11},
  pages = {115502},
  numpages = {5},
  year = {2015},
  month = {Sep},
  publisher = {American Physical Society},
  doi = {10.1103/PhysRevLett.115.115502},
  url = {https://doi.org/10.1103/PhysRevLett.115.115502}
}

@article{che2025,
  title={Magnetic order induced chiral phonons in a ferromagnetic {Weyl} semimetal},
  author = {Che, Mengqian and Liang, Jinxuan and Cui, Yunpeng and Li, Hao and Lu, Bingru and Sang, Wenbo and Li, Xiang and Dong, Xuebin and Zhao, Le and Zhang, Shuai and Sun, Tao and Jiang, Wanjun and Liu, Enke and Jin, Feng and Zhang, Tiantian and Yang, Luyi},
  journal={Phys. Rev. Lett.},
  volume={134},
  number={19},
  pages={196906},
  year={2025},
  publisher={APS},
 doi = {10.1103/PhysRevLett.134.196906},
  url = {https://doi.org/10.1103/PhysRevLett.134.196906}
}

@article{blochl1994,
  title={Projector augmented-wave method},
  author={Bl{\"o}chl, Peter E},
  journal={Phys. Rev. B},
  volume={50},
  number={24},
  pages={17953},
  year={1994},
  publisher={APS},
  doi={10.1103/PhysRevB.50.17953},
  url={https://doi.org/10.1103/PhysRevB.50.17953}
}

@article{may2022,
doi = {10.1088/2053-1583/ac34d9},
url = {https://doi.org/10.1088/2053-1583/ac34d9},
year = {2021},
month = {nov},
publisher = {IOP Publishing},
volume = {9},
number = {1},
pages = {015013},
author = {May, Andrew F and Yan, Jiaqiang and Hermann, Raphael and Du, Mao-Hua and McGuire, Michael A},
title = {Tuning the room temperature ferromagnetism in {Fe$_{5}$GeTe$_{2}$} by arsenic substitution},
journal = {2D Mater.}
}

@Article{ershadrad2022,
author={Ershadrad, Soheil and Ghosh, Sukanya
and Wang, Duo and Kvashnin, Yaroslav
and Sanyal, Biplab},
title={Unusual Magnetic Features in Two-Dimensional {Fe$_{5}$GeTe$_{2}$} Induced by Structural Reconstructions},
journal={J. Phys. Chem. Lett.},
year={2022},
month={Jun},
day={09},
publisher={American Chemical Society},
volume={13},
number={22},
pages={4877},
doi={10.1021/acs.jpclett.2c00692},
url={https://doi.org/10.1021/acs.jpclett.2c00692}
}

@Article{ghosh2023,
author={Ghosh, Sukanya
and Ershadrad, Soheil
and Borisov, Vladislav
and Sanyal, Biplab},
title={Unraveling effects of electron correlation in two-dimensional {Fe$_{n}$GeTe$_{2}$} ($n = 3$, 4, 5) by dynamical mean field theory},
journal={npj Comput. Mater.},
year={2023},
month={May},
day={29},
volume={9},
number={1},
pages={86},
issn={2057-3960},
doi={10.1038/s41524-023-01024-5},
url={https://doi.org/10.1038/s41524-023-01024-5}
}

@article{grimme2011,
author = {Grimme, Stefan and Ehrlich, Stephan and Goerigk, Lars},
title = {Effect of the damping function in dispersion corrected density functional theory},
journal = {J. Comput. Chem.},
volume = {32},
number = {7},
pages = {1456},
doi = {10.1002/jcc.21759},
url = {https://doi.org/10.1002/jcc.21759},
year = {2011}
}

@Article{wang2020,
author ="He, Junshan and Wang, Cong and Zhou, Bo and Zhao, Yu and Tao, Lili and Zhang, Han",
title  ="{2D} van der {Waals} heterostructures: processing, optical properties and applications in ultrafast photonics",
journal  ="Mater. Horiz.",
year  ="2020",
volume  ="7",
issue  ="11",
pages  ="2903",
publisher  ="The Royal Society of Chemistry",
doi  ="10.1039/D0MH00340A",
url  ="http://doi.org/10.1039/D0MH00340A"
}

@article{chen2013,
author = {Chen, Bin and Yang, Jin Hu and Wang, Hang Dong and Imai, Masaki and Ohta, Hiroto and Michioka, Chishiro and Yoshimura, Kazuyoshi and Fang, Ming Hu},
title = {Magnetic Properties of Layered Itinerant Electron Ferromagnet {Fe$_{3}$GeTe$_{2}$}},
journal = {J. Phys. Soc. Jpn.},
volume = {82},
number = {12},
pages = {124711},
year = {2013},
doi = {10.7566/JPSJ.82.124711},
URL = {https://doi.org/10.7566/JPSJ.82.124711}
}

@Article{jiang2022,
author={Jiang, Huaning
and Zang, Zhihao
and Wang, Xingguo
and Que, Haifeng
and Wang, Lei
and Si, Kunpeng
and Zhang, Peng
and Ye, Yu
and Gong, Yongji},
title={Thickness-Tunable Growth of Composition-Controllable Two-Dimensional {Fe$_{x}$GeTe$_{2}$}},
journal={Nano Lett.},
year={2022},
month={Dec},
day={14},
publisher={American Chemical Society},
volume={22},
number={23},
pages={9477},
issn={1530-6984},
doi={10.1021/acs.nanolett.2c03562},
url={https://doi.org/10.1021/acs.nanolett.2c03562}
}

@Article{kim2018,
author={Kim, Kyoo
and Seo, Junho
and Lee, Eunwoo
and Ko, K.-T.
and Kim, B. S.
and Jang, Bo Gyu
and Ok, Jong Mok
and Lee, Jinwon
and Jo, Youn Jung
and Kang, Woun
and Shim, Ji Hoon
and Kim, C.
and Yeom, Han Woong
and Il Min, Byung
and Yang, Bohm-Jung
and Kim, Jun Sung},
title={Large anomalous {Hall} current induced by topological nodal lines in a ferromagnetic van der {Waals} semimetal},
journal={Nat. Mater.},
year={2018},
month={Sep},
day={01},
volume={17},
number={9},
pages={794},
issn={1476-4660},
doi={10.1038/s41563-018-0132-3},
url={https://doi.org/10.1038/s41563-018-0132-3}
}

@article{kong2021,
author = {Kong, Xiangru and Berlijn, Tom and Liang, Liangbo},
title = {Thickness and Spin Dependence of {Raman} Modes in Magnetic Layered {Fe$_{3}$GeTe$_{2}$}},
journal = {Adv. Electron. Mater.},
volume = {7},
number = {7},
pages = {2001159},
doi = {10.1002/aelm.202001159},
url = {https://doi.org/10.1002/aelm.202001159},
year = {2021}
}

@article{kresse1996,
  title={Efficiency of ab-initio total energy calculations for metals and semiconductors using a plane-wave basis set},
  author={Kresse, Georg and Furthm{\"u}ller, J{\"u}rgen},
  journal={Comput. Mater. Sci.},
  volume={6},
  number={1},
  pages={15},
  year={1996},
  publisher={Elsevier},
  doi={10.1016/0927-0256(96)00008-0},
  url={https://doi.org/10.1016/0927-0256(96)00008-0}
}

@article{kresse1993,
  title={Ab initio molecular dynamics for liquid metals},
  author={Kresse, Georg and Hafner, J{\"u}rgen},
  journal={Phys. Rev. B},
  volume={47},
  number={1},
  pages={558},
  year={1993},
  publisher={APS},
  doi={10.1103/PhysRevB.47.558},
  url={https://doi.org/10.1103/PhysRevB.47.558}
}

@article{kresse1999,
  title = {From ultrasoft pseudopotentials to the projector augmented-wave method},
  author = {Kresse, G. and Joubert, D.},
  journal = {Phys. Rev. B},
  volume = {59},
  issue = {3},
  pages = {1758},
  numpages = {0},
  year = {1999},
  month = {Jan},
  publisher = {American Physical Society},
  doi = {10.1103/PhysRevB.59.1758},
  url = {http://doi.org/10.1103/PhysRevB.59.1758}
}

@article{li2018,
author = {Li, Qian and Yang, Mengmeng and Gong, Cheng and Chopdekar, Rajesh V. and N’Diaye, Alpha T. and Turner, John and Chen, Gong and Scholl, Andreas and Shafer, Padraic and Arenholz, Elke and Schmid, Andreas K. and Wang, Sheng and Liu, Kai and Gao, Nan and Admasu, Alemayehu S. and Cheong, Sang-Wook and Hwang, Chanyong and Li, Jia and Wang, Feng and Zhang, Xiang and Qiu, Ziqiang},
title = {Patterning-Induced Ferromagnetism of {Fe$_{3}$GeTe$_{2}$} van der {Waals} Materials beyond Room Temperature},
journal = {Nano Lett.},
volume = {18},
number = {9},
pages = {5974},
year = {2018},
doi = {10.1021/acs.nanolett.8b02806},
URL = {https://doi.org/10.1021/acs.nanolett.8b02806}
}

@article{li2025,
    author = {Li, Junbo and Wang, Yihao and Wu, Peng and Song, Jiangpeng and Li, Zhihao and Cao, Liang and Wang, Guopeng and Xiong, Yimin},
    title = {Spin–lattice coupling and anisotropic magneto-transport in {Fe}-ordered {Fe$_{5-x}$GeTe$_{2}$}},
    journal = {Appl. Phys. Lett.},
    volume = {127},
    number = {4},
    pages = {042401},
    year = {2025},
    month = {07},
    issn = {0003-6951},
    doi = {10.1063/5.0266796},
    url = {https://doi.org/10.1063/5.0266796}
}

@article{liechtenstein1995,
  title = {Density-functional theory and strong interactions: Orbital ordering in {Mott--Hubbard} insulators},
  author = {Liechtenstein, A. I. and Anisimov, V. I. and Zaanen, J.},
  journal = {Phys. Rev. B},
  volume = {52},
  issue = {8},
  pages = {R5467},
  numpages = {0},
  year = {1995},
  month = {Aug},
  publisher = {American Physical Society},
  doi = {10.1103/PhysRevB.52.R5467},
  url = {https://doi.org/10.1103/PhysRevB.52.R5467}
}

@article{ly2021,
author = {Ly, Trinh Thi and Park, Jungmin and Kim, Kyoo and Ahn, Hyo-Bin and Lee, Nyun Jong and Kim, Kwangsu and Park, Tae-Eon and Duvjir, Ganbat and Lam, Nguyen Huu and Jang, Kyuha and You, Chun-Yeol and Jo, Younghun and Kim, Se Kwon and Lee, Changgu and Kim, Sanghoon and Kim, Jungdae},
title = {Direct Observation of {Fe-Ge} Ordering in {Fe$_{5-x}$GeTe$_{2}$} Crystals and Resultant Helimagnetism},
journal = {Adv. Funct. Mater.},
volume = {31},
number = {17},
pages = {2009758},
doi = {10.1002/adfm.202009758},
url = {https://doi.org/10.1002/adfm.202009758},
year = {2021}
}

@article{may2016,
  title = {Magnetic structure and phase stability of the van der {Waals} bonded ferromagnet {Fe$_{3-x}$GeTe$_{2}$}},
  author = {May, Andrew F. and Calder, Stuart and Cantoni, Claudia and Cao, Huibo and McGuire, Michael A.},
  journal = {Phys. Rev. B},
  volume = {93},
  issue = {1},
  pages = {014411},
  numpages = {11},
  year = {2016},
  month = {Jan},
  publisher = {American Physical Society},
  doi = {10.1103/PhysRevB.93.014411},
  url = {https://doi.org/10.1103/PhysRevB.93.014411}
}

@article{may2020,
  title = {Tuning magnetic order in the van der {Waals} metal {Fe$_{5}$GeTe$_{2}$} by cobalt substitution},
  author = {May, Andrew F. and Du, Mao-Hua and Cooper, Valentino R. and McGuire, Michael A.},
  journal = {Phys. Rev. Mater.},
  volume = {4},
  issue = {7},
  pages = {074008},
  numpages = {9},
  year = {2020},
  month = {Jul},
  publisher = {American Physical Society},
  doi = {10.1103/PhysRevMaterials.4.074008},
  url = {https://doi.org/10.1103/PhysRevMaterials.4.074008}
}

@article{mi2023,
title = {Two-dimensional magnetic materials for spintronic devices},
journal = {Mater. Today Nano},
volume = {24},
pages = {100408},
year = {2023},
issn = {2588-8420},
doi = {10.1016/j.mtnano.2023.100408},
url = {https://doi.org/10.1016/j.mtnano.2023.100408},
author = {Mengjuan Mi and Han Xiao and Lixuan Yu and Yingxu Zhang and Yuanshuo Wang and Qiang Cao and Yilin Wang}
}

@article{milosavljevic2019,
  title = {Lattice dynamics and phase transitions in {Fe$_{3-x}$GeTe$_{2}$}},
  author = {Milosavljevi\'{c}, A. and \v{S}olaji\'{c}, A. and Djurdji\'{c}-Mijin, S. and Pe\v{s}i\'{c}, J. and Vi\v{s}i\'{c}, B. and Liu, Yu and Petrovic, C. and Lazarevi\'{c}, N. and Popovi\'{c}, Z. V.},
  journal = {Phys. Rev. B},
  volume = {99},
  issue = {21},
  pages = {214304},
  numpages = {7},
  year = {2019},
  month = {Jun},
  publisher = {American Physical Society},
  doi = {10.1103/PhysRevB.99.214304},
  url = {https://doi.org/10.1103/PhysRevB.99.214304}
}

@article{monkhorst1976,
  title={Special points for {Brillouin-zone} integrations},
  author={Monkhorst, Hendrik J and Pack, James D},
  journal={Phys. Rev. B},
  volume={13},
  number={12},
  pages={5188},
  year={1976},
  publisher={APS},
  doi = {10.1103/PhysRevB.13.5188},
  url = {https://doi.org/10.1103/PhysRevB.13.5188}
}

@Article{moon2024,
author={Moon, Alex
and Li, Yue
and McKeever, Conor
and Casas, Brian W.
and Bravo, Moises
and Zheng, Wenkai
and Macy, Juan
and Petford-Long, Amanda K.
and McCandless, Gregory T.
and Chan, Julia Y.
and Phatak, Charudatta
and Santos, Elton J. G.
and Balicas, Luis},
title={Writing and Detecting Topological Charges in Exfoliated {Fe$_{5-x}$GeTe$_{2}$}},
journal={ACS Nano},
year={2024},
month={Feb},
day={06},
publisher={American Chemical Society},
volume={18},
number={5},
pages={4216},
issn={1936-0851},
doi={10.1021/acsnano.3c09234},
url={https://doi.org/10.1021/acsnano.3c09234}
}

@article{perdew1996,
  title={Generalized gradient approximation made simple},
  author={Perdew, John P and Burke, Kieron and Ernzerhof, Matthias},
  journal={Phys. Rev. Lett.},
  volume={77},
  number={18},
  pages={3865},
  year={1996},
  publisher={APS},
  doi = {10.1103/PhysRevLett.77.3865},
  url = {https://doi.org/10.1103/PhysRevLett.77.3865}
}

@Article{ren2023,
AUTHOR = {Ren, Hongtao and Lan, Mu},
TITLE = {Progress and Prospects in Metallic {Fe$_{x}$GeTe$_{2}$} ($3 \leq x \leq 7$) Ferromagnets},
JOURNAL = {Molecules},
VOLUME = {28},
YEAR = {2023},
NUMBER = {21},
pages = {7244},
URL = {https://doi.org/10.3390/molecules28217244},
PubMedID = {37959664},
ISSN = {1420-3049},
DOI = {10.3390/molecules28217244}
}

@article{seo2020,
author = {Junho Seo  and Duck Young Kim  and Eun Su An  and Kyoo Kim  and Gi-Yeop Kim  and Soo-Yoon Hwang  and Dong Wook Kim  and Bo Gyu Jang  and Heejung Kim  and Gyeongsik Eom  and Seung Young Seo  and Roland Stania  and Matthias Muntwiler  and Jinwon Lee  and Kenji Watanabe  and Takashi Taniguchi  and Youn Jung Jo  and Jieun Lee  and Byung Il Min  and Moon Ho Jo  and Han Woong Yeom  and Si-Young Choi  and Ji Hoon Shim  and Jun Sung Kim },
title = {Nearly room temperature ferromagnetism in a magnetic metal-rich van der {Waals} metal},
journal = {Sci. Adv.},
volume = {6},
number = {3},
pages = {eaay8912},
year = {2020},
doi = {10.1126/sciadv.aay8912},
URL = {https://doi.org/10.1126/sciadv.aay8912}
}

@Article{sierra2021,
author={Sierra, Juan F.
and Fabian, Jaroslav
and Kawakami, Roland K.
and Roche, Stephan
and Valenzuela, Sergio O.},
title={Van der {Waals} heterostructures for spintronics and opto-spintronics},
journal={Nat. Nanotechnol.},
year={2021},
month={Aug},
day={01},
volume={16},
number={8},
pages={856},
issn={1748-3395},
doi={10.1038/s41565-021-00936-x},
url={https://doi.org/10.1038/s41565-021-00936-x}
}

@article{stahl2018,
author = {Stahl, Juliane and Shlaen, Evgeniya and Johrendt, Dirk},
title = {The van der Waals Ferromagnets {Fe$_{5-\delta}$GeTe$_{2}$} and {Fe$_{5-\delta-x}$Ni$_{x}$GeTe$_{2}$} -- Crystal Structure, Stacking Faults, and Magnetic Properties},
journal = {Z. Anorg. Allg. Chem.},
volume = {644},
number = {24},
pages = {1923},
doi = {10.1002/zaac.201800456},
url = {https://doi.org/10.1002/zaac.201800456},
year = {2018}
}

@article{tadano2014,
doi = {10.1088/0953-8984/26/22/225402},
url = {https://doi.org/10.1088/0953-8984/26/22/225402},
year = {2014},
month = {may},
publisher = {IOP Publishing},
volume = {26},
number = {22},
pages = {225402},
author = {Tadano, T. and Gohda, Y. and Tsuneyuki, S.},
title = {Anharmonic force constants extracted from first-principles molecular dynamics: applications to heat transfer simulations},
journal = {J. Phys. Condens. Matter}
}

@article{togo2023a,
  author  = {Togo, Atsushi},
  title   = {First-principles Phonon Calculations with Phonopy and Phono3py},
  journal = {J. Phys. Soc. Jpn.},
  volume  = {92},
  number  = {1},
  pages   = {012001},
  year    = {2023},
  doi     = {10.7566/JPSJ.92.012001},
url = {http://doi.org/10.7566/JPSJ.92.012001}
}

@article{togo2023b,
  author  = {Togo, Atsushi and Chaput, Laurent and Tadano, Terumasa and Tanaka, Isao},
  title   = {Implementation strategies in phonopy and phono3py},
  journal = {J. Phys. Condens. Matter},
  volume  = {35},
  number  = {35},
  pages   = {353001},
  year    = {2023},
  doi     = {10.1088/1361-648X/acd831},
url = {http://doi.org/10.1088/1361-648X/acd831}
}

@article{zhu2016,
  title = {Electronic correlation and magnetism in the ferromagnetic metal {Fe$_{3}$GeTe$_{2}$}},
  author = {Zhu, Jian-Xin and Janoschek, Marc and Chaves, D. S. and Cezar, J. C. and Durakiewicz, Tomasz and Ronning, Filip and Sassa, Yasmine and Mansson, Martin and Scott, B. L. and Wakeham, N. and Bauer, Eric D. and Thompson, J. D.},
  journal = {Phys. Rev. B},
  volume = {93},
  issue = {14},
  pages = {144404},
  numpages = {6},
  year = {2016},
  month = {Apr},
  publisher = {American Physical Society},
  doi = {10.1103/PhysRevB.93.144404},
  url = {https://doi.org/10.1103/PhysRevB.93.144404}
}

@article{adhikari2025,
  title={Room Temperature Evolution of Laser-Induced Ultrafast Spin and Phonon Dynamics in {2D} van der {Waals} Magnets {Fe$_x$GeTe$_2$} ($x= 3, 4, 5$)},
  author={Adhikari, Arundhati and Mahato, Bipul Kumar and Sahoo, Sourav and Mukhopadhyay, Suchetana and Palit, Mainak and Bera, Satyabrata and Datta, Subhadeep and Mondal, Mintu and Barman, Anjan},
  journal={Adv. Funct. Mater.},
  volume={35},
  number={13},
  pages={2418006},
  year={2025},
  publisher={Wiley Online Library},
DOI={10.1002/adfm.202418006},
URL={https://doi.org/10.1002/adfm.202418006}
}

@article{may2019a,
  title={Ferromagnetism near room temperature in the cleavable van der {Waals} crystal {Fe$_5$GeTe$_2$}},
  author={May, Andrew F and Ovchinnikov, Dmitry and Zheng, Qiang and Hermann, Raphael and Calder, Stuart and Huang, Bevin and Fei, Zaiyao and Liu, Yaohua and Xu, Xiaodong and McGuire, Michael A},
  journal={ACS Nano.},
  volume={13},
  number={4},
  pages={4436},
  year={2019},
  publisher={ACS Publications},
doi = {10.1021/acsnano.8b09660},
URL = {https://doi.org/10.1021/acsnano.8b09660}
}

@article{zhang2020c,
  title={Itinerant ferromagnetism in van der {Waals} {Fe$_{5-x}$GeTe$_2$} crystals above room temperature},
  author = {Zhang, Hongrui and Chen, Rui and Zhai, Kun and Chen, Xiang and Caretta, Lucas and Huang, Xiaoxi and Chopdekar, Rajesh V. and Cao, Jinhua and Sun, Jirong and Yao, Jie and Birgeneau, Robert and Ramesh, Ramamoorthy},
  journal={Phys. Rev. B},
  volume={102},
  number={6},
  pages={064417},
  year={2020},
  publisher={APS},
doi = {10.1103/PhysRevB.102.064417},
  url = {https://doi.org/10.1103/PhysRevB.102.064417}
}

@article{may2019b,
  title={Physical properties and thermal stability of {Fe$_{5-x}$GeTe$_2$} single crystals},
  author={May, Andrew F and Bridges, Craig A and McGuire, Michael A},
  journal={Phys. Rev. Mater.},
  volume={3},
  number={10},
  pages={104401},
  year={2019},
  publisher={APS},
  doi = {10.1103/PhysRevMaterials.3.104401},
  url = {https://doi.org/10.1103/PhysRevMaterials.3.104401}
}

@article{liu2022,
  title={Layer-dependent magnetic phase diagram in {Fe$_n$ GeTe$_2$} (3$ \leq$ n $ \leq$ 7) ultrathin films},
  author={Liu, Qinxi and Xing, Jianpei and Jiang, Zhou and Guo, Yu and Jiang, Xue and Qi, Yan and Zhao, Jijun},
  journal={Commun. Phys.},
  volume={5},
  number={1},
  pages={140},
  year={2022},
  publisher={Nature Publishing Group UK London},
URL = {https://doi.org/10.1038/s42005-022-00921-3},
DOI  ={10.1038/s42005-022-00921-3} 

}

@article{han2022,
  title={Complex \text{Raman} tensor in helicity-changing \text{Raman} spectra of black phosphorus under circularly polarized light},
  author={Han, Shiyi and Zhao, Yan and Tuan Hung, Nguyen and Xu, Bo and Saito, Riichiro and Zhang, Jin and Tong, Lianming},
  journal={J. Phys. Chem. Lett.},
  volume={13},
  number={5},
  pages={1241},
  year={2022},
  publisher={ACS Publications},
DOI={10.1021/acs.jpclett.1c03826}, 
URL={https://doi.org/10.1021/acs.jpclett.1c03826}   

}

@article{pimenta2021,
  title={Polarized {Raman} spectroscopy in low-symmetry {2D} materials: angle-resolved experiments and complex number tensor elements},
  author={Pimenta, Marcos A and Resende, Geovani C and Ribeiro, Henrique B and Carvalho, Bruno R},
  journal={Phys. Chem. Chem. Phys.},
  volume={23},
  number={48},
  pages={27103},
  year={2021},
  publisher={Royal Society of Chemistry},
  DOI={10.1039/D1CP03626B},
  URL={https://doi.org/10.1039/D1CP03626B}
}

@article{oishi2024,
  title={Selective observation of enantiomeric chiral phonons in $\alpha$-quartz},
  author={Oishi, Eiichi and Fujii, Yasuhiro and Koreeda, Akitoshi},
  journal={Phys. Rev. B},
  volume={109},
  number={10},
  pages={104306},
  year={2024},
  publisher={APS},
 doi = {10.1103/PhysRevB.109.104306},
  url = {https://doi.org/10.1103/PhysRevB.109.104306}
}

@article{cracknell1969,
  title={Scattering matrices for the {Raman} effect in magnetic crystals},
  author={Cracknell, A. P.},
  journal={J. Phys. C: Solid State Phys.},
  volume={2},
  number={3},
  pages={500},
  year={1969},
  publisher={IOP Publishing},
  DOI={10.1088/0022-3719/2/3/314},
  URL={https://doi.org/10.1088/0022-3719/2/3/314/meta}
}

@article{momma.izumi.11,
author = "Momma, K. and Izumi, F.",
title = "{{\sc vesta3} for three-dimensional visualization of crystal, volumetric and morphology data}",
journal = "J. Appl. Crystallogr.",
year = "2011",
volume = "44",
number = "6",
pages = "1272",
month = "Dec",
doi = {10.1107/S0021889811038970},
url = {https://doi.org/10.1107/S0021889811038970}
}

@article{yang2021a,
author = {Yang, Shengxue and Zhang, Tianle and Jiang, Chengbao},
title = {van der {Waals} Magnets: Material Family, Detection and Modulation of Magnetism, and Perspective in Spintronics},
journal = {Adv. Sci.},
volume = {8},
number = {2},
pages = {2002488},
doi = {10.1002/advs.202002488},
url = {https://doi.org/10.1002/advs.202002488},
year = {2021}
}

@article{deiseroth2006,
author = {Deiseroth, Hans-Jörg and Aleksandrov, Krasimir and Reiner, Christof and Kienle, Lorenz and Kremer, Reinhard K.},
title = {{Fe$_{3}$GeTe$_{2}$} and {Ni$_{3}$GeTe$_{2}$} -- Two New Layered Transition-Metal Compounds: Crystal Structures, {HRTEM} Investigations, and Magnetic and Electrical Properties},
journal = {Eur. J. Inorg. Chem.},
volume = {2006},
number = {8},
pages = {1561},
url = {https://doi.org/10.1002/ejic.200501020},
doi = {10.1002/ejic.200501020},
year = {2006}
}

@Article{wang.lu.23,
author={Wang, Hangtian
and Lu, Haichang
and Guo, Zongxia
and Li, Ang
and Wu, Peichen
and Li, Jing
and Xie, Weiran
and Sun, Zhimei
and Li, Peng
and Damas, H{\'e}lo{\"i}se
and Friedel, Anna Maria
and Migot, Sylvie
and Ghanbaja, Jaafar
and Moreau, Luc
and Fagot-Revurat, Yannick
and Petit-Watelot, S{\'e}bastien
and Hauet, Thomas
and Robertson, John
and Mangin, St{\'e}phane
and Zhao, Weisheng
and Nie, Tianxiao},
title={Interfacial engineering of ferromagnetism in wafer-scale van der {Waals} {Fe$_{4}$GeTe$_{2}$} far above room temperature},
journal={Nat. Commun.},
year={2023},
month={Apr},
day={29},
volume={14},
number={1},
pages={2483},
issn={2041-1723},
doi={10.1038/s41467-023-37917-8},
url={https://doi.org/10.1038/s41467-023-37917-8}
}

@article{tian2020,
    author = {Tian, Congkuan and Pan, Feihao and Xu, Sheng and Ai, Kun and Xia, Tianlong and Cheng, Peng},
    title = {Tunable magnetic properties in van der {Waals} crystals {(Fe$_{1-x}$Co$_{x}$)$_{5}$GeTe$_{2}$}},
    journal = {Appl. Phys. Lett.},
    volume = {116},
    number = {20},
    pages = {202402},
    year = {2020},
    month = {05},
    issn = {0003-6951},
    doi = {10.1063/5.0006337},
    url = {https://doi.org/10.1063/5.0006337}
}

@article{zhang2024,
    author = {Zhang, Junchao and Wang, Ziwen and Xing, Yu and Luo, Xiong and Wang, Zhicheng and Wang, Guopeng and Shen, Aoli and Ye, Haoran and Dong, Shuai and Li, Linglong},
    title = {Enhanced magnetic and electrical properties of {Co}-doped {Fe$_{5}$GeTe$_{2}$}},
    journal = {Appl. Phys. Lett.},
    volume = {124},
    number = {10},
    pages = {103103},
    year = {2024},
    month = {03},
    issn = {0003-6951},
    doi = {10.1063/5.0194813},
    url = {https://doi.org/10.1063/5.0194813}
}

@Article{birch2024,
author={Birch, Max T. and Yasin, Fehmi S.
and Litzius, Kai and Powalla, Lukas 
and Wintz, Sebastian and Schulz, Frank
and Kossak, Alexander  E. and Weigand, Markus
and Scholz, Tanja and Lotsch, Bettina V.
and Sch{\"u}tz, Gisela and Yu, Xiuzhen Z.
and Burghard, Marko},
title={Influence of Magnetic Sublattice Ordering on Skyrmion Bubble Stability in {2D} Magnet {Fe$_{5}$GeTe$_{2}$}},
journal={ACS Nano},
year={2024},
month={Jul},
day={16},
publisher={American Chemical Society},
volume={18},
number={28},
pages={18246},
issn={1936-0851},
doi={10.1021/acsnano.4c00853},
url={https://doi.org/10.1021/acsnano.4c00853}
}

@Article{schmitt2022,
author={Schmitt, Maurice and Denneulin, Thibaud
and Kov{\'a}cs, Andr{\'a}s and Saunderson, Tom G.
and R{\"u}{\ss}mann, Philipp and Shahee, Aga
and Scholz, Tanja and Tavabi, Amir H.
and Gradhand, Martin and Mavropoulos, Phivos
and Lotsch, Bettina V. and Dunin-Borkowski, Rafal E.
and Mokrousov, Yuriy and Bl{\"u}gel, Stefan
and Kl{\"a}ui, Mathias},
title={Skyrmionic spin structures in layered {Fe$_5$GeTe$_2$} up to room temperature},
journal={Commun. Phys.},
year={2022},
month={Oct},
day={18},
volume={5},
number={1},
pages={254},
issn={2399-3650},
doi={10.1038/s42005-022-01031-w},
url={https://doi.org/10.1038/s42005-022-01031-w}
}

@Article{bae2022,
author={Bae, Soungmin and Matsumoto, Kana
and Raebiger, Hannes and Shudo, Ken-ichi
and Kim, Yong-Hoon and Handeg{\aa}rd, {\O}rjan Sele
and Nagao, Tadaaki and Kitajima, Masahiro
and Sakai, Yuji and Zhang, Xiang
and Vajtai, Robert and Ajayan, Pulickel
and Kono, Junichiro and Takeda, Jun
and Katayama, Ikufumi},
title={K-point longitudinal acoustic phonons are responsible for ultrafast intervalley scattering in monolayer {MoSe$_{2}$}},
journal={Nat. Commun.},
year={2022},
month={Jul},
day={25},
volume={13},
number={1},
pages={4279},
issn={2041-1723},
doi={10.1038/s41467-022-32008-6},
url={https://doi.org/10.1038/s41467-022-32008-6}
}

@article{nika2009,
  title = {Phonon thermal conduction in graphene: Role of Umklapp and edge roughness scattering},
  author = {Nika, D. L. and Pokatilov, E. P. and Askerov, A. S. and Balandin, A. A.},
  journal = {Phys. Rev. B},
  volume = {79},
  issue = {15},
  pages = {155413},
  numpages = {12},
  year = {2009},
  month = {Apr},
  publisher = {American Physical Society},
  doi = {10.1103/PhysRevB.79.155413},
  url = {https://doi.org/10.1103/PhysRevB.79.155413}
}

@Article{wan2021,
author ="Wan, Yi and Cheng, Xing and Li, Yanfang and Wang, Yaqian and Du, Yongping and Zhao, Yibin and Peng, Bo and Dai, Lun and Kan, Erjun",
title  ="Manipulating the {Raman} scattering rotation via magnetic field in an {MoS$_{2}$} monolayer",
journal  ="RSC Adv.",
year  ="2021",
volume  ="11",
issue  ="7",
pages  ="4035",
publisher  ="The Royal Society of Chemistry",
doi  ="10.1039/D0RA09350E",
url  ="http://dx.doi.org/10.1039/D0RA09350E"
}

@article{libbi2021,
doi = {10.1088/2053-1583/abc5ce},
url = {https://doi.org/10.1088/2053-1583/abc5ce},
year = {2020},
month = {dec},
publisher = {IOP Publishing},
volume = {8},
number = {1},
pages = {015026},
author = {Libbi, Francesco and Bonini, Nicola and Marzari, Nicola},
title = {Thermomechanical properties of honeycomb lattices from internal-coordinates potentials: the case of graphene and hexagonal boron nitride},
journal = {2D Mater.}
}

@article{saito2016,
doi = {10.1088/0953-8984/28/35/353002},
url = {https://doi.org/10.1088/0953-8984/28/35/353002},
year = {2016},
month = {jul},
publisher = {IOP Publishing},
volume = {28},
number = {35},
pages = {353002},
author = {Saito, R and Tatsumi, Y and Huang, S and Ling, X and Dresselhaus, M S},
title = {Raman spectroscopy of transition metal dichalcogenides},
journal = {J. Phys.: Condens. Matter}
}

@article{mobaraki2019,
  title = {Temperature-dependent phonon spectrum of transition metal dichalcogenides calculated from the spectral energy density: Lattice thermal conductivity as an application},
  author = {Mobaraki, Arash and Sevik, Cem and Yapicioglu, Haluk and Cakir, Deniz and Gulseren, Oguz},
  journal = {Phys. Rev. B},
  volume = {100},
  issue = {3},
  pages = {035402},
  numpages = {6},
  year = {2019},
  month = {Jul},
  publisher = {American Physical Society},
  doi = {10.1103/PhysRevB.100.035402},
  url = {https://doi.org/10.1103/PhysRevB.100.035402}
}

@article{wilczynski2022,
title = {Phonon anharmonicity in multi-layered {WS$_{2}$} explored by first-principles and {Raman} studies},
journal = {Acta Materialia},
volume = {240},
pages = {118299},
year = {2022},
issn = {1359-6454},
doi = {10.1016/j.actamat.2022.118299},
url = {https://doi.org/10.1016/j.actamat.2022.118299},
author = {Konrad Wilczy\'{n}ski and Arkadiusz P. Gertych and Karolina Czerniak-\L{}osiewicz and Jakub Sitek and Mariusz Zdrojek}
}

@article{ishito2023,
author = {Ishito, Kyosuke and Mao, Huiling and Kobayashi, Kaya and Kousaka, Yusuke and Togawa, Yoshihiko and Kusunose, Hiroaki and Kishine, Jun-ichiro and Satoh, Takuya},
title = {Chiral phonons: circularly polarized {Raman} spectroscopy and ab initio calculations in a chiral crystal tellurium},
journal = {Chirality},
volume = {35},
number = {6},
pages = {338},
doi = {10.1002/chir.23544},
url = {https://doi.org/10.1002/chir.23544},
year = {2023}
}

@Article{gish2024,
author={Gish, J. Tyler
and Lebedev, Dmitry
and Song, Thomas W.
and Sangwan, Vinod K.
and Hersam, Mark C.},
title={Van der {Waals} opto-spintronics},
journal={Nat. Electron.},
year={2024},
month={May},
day={01},
volume={7},
number={5},
pages={336},
issn={2520-1131},
doi={10.1038/s41928-024-01167-3},
url={https://doi.org/10.1038/s41928-024-01167-3}
}

@Article{zhao2025,
author={Zhao, Zhiyuan
and Lin, Yijie
and Avsar, Ahmet},
title={Novel spintronic effects in two-dimensional van der {Waals} heterostructures},
journal={npj 2D Mater. Appl.},
year={2025},
month={Apr},
day={09},
volume={9},
number={1},
pages={30},
issn={2397-7132},
doi={10.1038/s41699-025-00546-4},
url={https://doi.org/10.1038/s41699-025-00546-4}
}

@Article{kajale2024,
author={Kajale, Shivam N.
and Nguyen, Thanh
and Chao, Corson A.
and Bono, David C.
and Boonkird, Artittaya
and Li, Mingda
and Sarkar, Deblina},
title={Current-induced switching of a van der {Waals} ferromagnet at room temperature},
journal={Nat. Commun.},
year={2024},
month={Feb},
day={19},
volume={15},
number={1},
pages={1485},
issn={2041-1723},
doi={10.1038/s41467-024-45586-4},
url={https://doi.org/10.1038/s41467-024-45586-4}
}

@Article{zhang2025,
author={Zhang, Zhiheng
and Sun, Rong
and Wang, Zhongchang},
title={Recent Advances in Two-Dimensional Ferromagnetic Materials-Based van der {Waals} Heterostructures},
journal={ACS Nano},
year={2025},
month={Jan},
day={14},
publisher={American Chemical Society},
volume={19},
number={1},
pages={187},
issn={1936-0851},
doi={10.1021/acsnano.4c14733},
url={https://doi.org/10.1021/acsnano.4c14733}
}

@Article{geim2013,
author={Geim, A. K.
and Grigorieva, I. V.},
title={Van der {Waals} heterostructures},
journal={Nature},
year={2013},
month={Jul},
day={01},
volume={499},
number={7459},
pages={419},
issn={1476-4687},
doi={10.1038/nature12385},
url={https://doi.org/10.1038/nature12385}
}

@article{novoselov2016,
author = {K. S. Novoselov  and A. Mishchenko  and A. Carvalho  and A. H. Castro Neto },
title = {2D materials and van der {Waals} heterostructures},
journal = {Science},
volume = {353},
number = {6298},
pages = {aac9439},
year = {2016},
doi = {10.1126/science.aac9439},
URL = {https://doi.org/10.1126/science.aac9439}
}

@article{zhong2017,
author = {Ding Zhong  and Kyle L. Seyler  and Xiayu Linpeng  and Ran Cheng  and Nikhil Sivadas  and Bevin Huang  and Emma Schmidgall  and Takashi Taniguchi  and Kenji Watanabe  and Michael A. McGuire  and Wang Yao  and Di Xiao  and Kai-Mei C. Fu  and Xiaodong Xu },
title = {Van der {Waals} engineering of ferromagnetic semiconductor heterostructures for spin and valleytronics},
journal = {Sci. Adv.},
volume = {3},
number = {5},
pages = {e1603113},
year = {2017},
doi = {10.1126/sciadv.1603113},
URL = {https://doi.org/10.1126/sciadv.1603113}
}

@Article{zollner2025,
author={Zollner, Klaus
and Kurpas, Marcin
and Gmitra, Martin
and Fabian, Jaroslav},
title={First-principles determination of spin--orbit coupling parameters in two-dimensional materials},
journal={Nat. Rev. Phys.},
year={2025},
month={May},
day={01},
volume={7},
number={5},
pages={255},
issn={2522-5820},
doi={10.1038/s42254-025-00818-4},
url={https://doi.org/10.1038/s42254-025-00818-4}
}

@article{xu2024,
author = {Xu, Hang and Xue, Yue and Liu, Zhenqi and Tang, Qing and Wang, Tianyi and Gao, Xichan and Qi, Yaping and Chen, Yong P. and Ma, Chunlan and Jiang, Yucheng},
title = {Van der {Waals} Heterostructures for Photoelectric, Memory, and Neural Network Applications},
journal = {Small Sci},
volume = {4},
number = {4},
pages = {2300213},
doi = {10.1002/smsc.202300213},
url = {https://doi.org/10.1002/smsc.202300213},
year = {2024}
}

@article{monda2021,
  title = {Critical behavior in the van der {Waals} itinerant ferromagnet {Fe$_{4}$GeTe$_{2}$}},
  author = {Mondal, Suchanda and Khan, Nazir and Mishra, Smruti Medha and Satpati, Biswarup and Mandal, Prabhat},
  journal = {Phys. Rev. B},
  volume = {104},
  issue = {9},
  pages = {094405},
  numpages = {9},
  year = {2021},
  month = {Sep},
  publisher = {American Physical Society},
  doi = {10.1103/PhysRevB.104.094405},
  url = {https://doi.org/10.1103/PhysRevB.104.094405}
}

@article{wang2025,
    author = {Wang, Xiaocui and Li, Peiling and Li, Yongkai and Yang, Xue and Lyu, Zhaozheng and Qu, Fanming and Shen, Jie and Jing, Xiunian and Liu, Guangtong and Lu, Li and Duan, Junxi and Wang, Zhiwei},
    title = {Signature of canted ferromagnetism in van der {Waals} {Fe$_{5-x}$GeTe$_{2}$} flakes},
    journal = {Appl. Phys. Lett.},
    volume = {126},
    number = {26},
    pages = {262405},
    year = {2025},
    month = {07},
    issn = {0003-6951},
    doi = {10.1063/5.0253974},
    url = {https://doi.org/10.1063/5.0253974}
}

@Article{zhao2023,
author={Zhao, Ruijie and Wu, Yanfei
and Yan, Shaohua and Liu, Xinjie
and Huang, He and Gao, Yang
and Zhu, Mengyuan and Shen, Jianxin
and Shen, Shipeng and Xu, Weifeng
and Zhang, Zeyu and Zhang, Liyuan
and Zhang, Jingyan and Zheng, Xinqi
and Lei, Hechang and Zhang, Ying
and Wang, Shouguo},
title={Magnetoresistance anomaly in {Fe$_{5}$GeTe$_{2}$} homo-junctions induced by its intrinsic transition},
journal={Nano Res.},
year={2023},
month={Jul},
day={01},
volume={16},
number={7},
pages={10443},
issn={1998-0000},
doi={10.1007/s12274-023-5609-y},
url={https://doi.org/10.1007/s12274-023-5609-y}
}

@article{suzuki2023,
    author = {Suzuki, Ryuki and Gao, Tenghua and Nakayama, Hiroki and Ando, Kazuya},
    title = {Extrinsic anomalous {Hall} effect in van der {Waals} ferromagnet {Fe$_{5}$GeTe$_{2}$}},
    journal = {AIP Advances},
    volume = {13},
    number = {5},
    pages = {055311},
    year = {2023},
    month = {05},
    issn = {2158-3226},
    doi = {10.1063/5.0112456},
    url = {https://doi.org/10.1063/5.0112456}
}

@article{dang2023,
author = {Dang, Ngoc-Toan and Kozlenko, Denis P. and Lis, Olga N. and Kichanov, Sergey E. and Lukin, Yevgenii V. and Golosova, Natalia O. and Savenko, Boris N. and Duong, Dinh-Loc and Phan, The-Long and Tran, Tuan-Anh and Phan, Manh-Huong},
title = {High Pressure-Driven Magnetic Disorder and Structural Transformation in {Fe$_{3}$GeTe$_{2}$}: Emergence of a Magnetic Quantum Critical Point},
journal = {Adv. Sci.},
volume = {10},
number = {9},
pages = {2206842},
doi = {10.1002/advs.202206842},
url = {https://doi.org/10.1002/advs.202206842},
year = {2023}
}

@Article{xu2020b,
title = {Possible Tricritical Behavior and Anomalous Lattice Softening in van der {Waals} Itinerant Ferromagnet {Fe$_{3}$GeTe$_{2}$} under High Pressure},
journal = {Chin. Phys. Lett.},
volume = {37},
number = {7},
pages = {076202},
year = {2020},
issn = {},
doi = {10.1088/0256-307X/37/7/076202},	
url = {http://doi.org/10.1088/0256-307X/37/7/076202},
author = {Jie-Min Xu and Shu-Yang Wang and Wen-Jun Wang and Yong-Hui Zhou and Xu-Liang Chen and Zhao-Rong Yang and Zhe Qu}
}

@Article{zhang2025b,
author={Zhang, Gaojie
and Wu, Hao
and Yang, Li
and Jin, Wen
and Xiao, Bichen
and Zhang, Wenfeng
and Chang, Haixin},
title={Lattice Vibration, Raman Modes and Room-Temperature Spin--Phonon Coupling in Intrinsic Two-Dimensional van der {Waals} Ferromagnetic {Fe$_{3}$GaTe$_{2}$}},
journal={ACS Materials Lett.},
year={2025},
month={Apr},
day={07},
publisher={American Chemical Society},
volume={7},
number={4},
pages={1289},
doi={10.1021/acsmaterialslett.4c02526},
url={https://doi.org/10.1021/acsmaterialslett.4c02526}
}

@article{hung.2024,
title = {{QERaman}: An open-source program for calculating resonance {Raman} spectra based on {Quantum ESPRESSO}},
journal = {	Comput. Phys. Commun.},
volume = {295},
pages = {108967},
year = {2024},
issn = {0010-4655},
doi = {10.1016/j.cpc.2023.108967},
url = {https://doi.org/10.1016/j.cpc.2023.108967},
author = {Nguyen Tuan Hung and Jianqi Huang and Yuki Tatsumi and Teng Yang and Riichiro Saito}
}

@article{giannozzi.20,
    author = {Giannozzi, Paolo and Baseggio, Oscar and Bonfà, Pietro and Brunato, Davide and Car, Roberto and Carnimeo, Ivan and Cavazzoni, Carlo and de Gironcoli, Stefano and Delugas, Pietro and Ferrari Ruffino, Fabrizio and Ferretti, Andrea and Marzari, Nicola and Timrov, Iurii and Urru, Andrea and Baroni, Stefano},
    title = {{Quantum ESPRESSO} toward the exascale},
    journal = {J. Chem. Phys.},
    volume = {152},
    number = {15},
    pages = {154105},
    year = {2020},
    month = {04},
    issn = {0021-9606},
    doi = {10.1063/5.0005082},
    url = {https://doi.org/10.1063/5.0005082}
}

@Article{wang2024,
author={Wang, Tingting
and Sun, Hong
and Li, Xiaozhe
and Zhang, Lifa},
title={Chiral Phonons: Prediction, Verification, and Application},
journal={Nano Lett.},
year={2024},
month={Apr},
day={17},
publisher={American Chemical Society},
volume={24},
number={15},
pages={4311},
issn={1530-6984},
doi={10.1021/acs.nanolett.4c00606},
url={https://doi.org/10.1021/acs.nanolett.4c00606}
}

@Article{lin2019,
author={Lin, Miao-Ling
and Zhou, Yu
and Wu, Jiang-Bin
and Cong, Xin
and Liu, Xue-Lu
and Zhang, Jun
and Li, Hai
and Yao, Wang
and Tan, Ping-Heng},
title={Cross-dimensional electron--phonon coupling in van der {Waals} heterostructures},
journal={Nature Commun.},
year={2019},
month={Jun},
day={03},
volume={10},
number={1},
pages={2419},
issn={2041-1723},
doi={10.1038/s41467-019-10400-z},
url={https://doi.org/10.1038/s41467-019-10400-z}
}

@article{resende2021,
doi = {10.1088/2053-1583/abce07},
url = {https://doi.org/10.1088/2053-1583/abce07},
year = {2020},
month = {dec},
publisher = {IOP Publishing},
volume = {8},
number = {2},
pages = {025002},
author = {Resende, Geovani C and Ribeiro, Guilherme A S and Silveira, Orlando J and Lemos, Jessica S and Brant, Juliana C and Rhodes, Daniel and Balicas, Luis and Terrones, Mauricio and Mazzoni, Mario S C and Fantini, Cristiano and Carvalho, Bruno R and Pimenta, Marcos A},
title = {Origin of the complex {Raman} tensor elements in single-layer triclinic {ReSe$_{2}$}},
journal = {2D Mater.}
}

@Article{sun2025,
author ="Sun, Yue and Liu, Bo and Gan, Chee Kwan and Xia, Shian and Lin, Haoyun and Liu, Sheng and Yu, Ting",
title  ="Strong and reciprocal magneto-phonon effects in a {2D} antiferromagnetic semiconductor {FePSe$_{3}$}",
journal  ="Nanoscale",
year  ="2025",
volume  ="17",
issue  ="14",
pages  ="8476",
publisher  ="The Royal Society of Chemistry",
doi  ="10.1039/D4NR04743E",
url  ="http://dx.doi.org/10.1039/D4NR04743E"
}

@article{liu2020,
author = {Zhen Liu  and Kai Guo  and Guangwei Hu  and Zhongtai Shi  and Yue Li  and Linbo Zhang  and Haiyan Chen  and Li Zhang  and Peiheng Zhou  and Haipeng Lu  and Miao-Ling Lin  and Sizhao Liu  and Yingchun Cheng  and Xue Lu Liu  and Jianliang Xie  and Lei Bi  and Ping-Heng Tan  and Longjiang Deng  and Cheng-Wei Qiu  and Bo Peng },
title = {Observation of nonreciprocal magnetophonon effect in nonencapsulated few-layered {CrI$_{3}$}},
journal = {Sci. Adv.},
volume = {6},
number = {43},
pages = {eabc7628},
year = {2020},
doi = {10.1126/sciadv.abc7628},
URL = {https://www.science.org/doi/abs/10.1126/sciadv.abc7628}
}

@Article{mccreary2020,
author={McCreary, Amber
and Mai, Thuc T.
and Utermohlen, Franz G.
and Simpson, Jeffrey R.
and Garrity, Kevin F.
and Feng, Xiaozhou
and Shcherbakov, Dmitry
and Zhu, Yanglin
and Hu, Jin
and Weber, Daniel
and Watanabe, Kenji
and Taniguchi, Takashi
and Goldberger, Joshua E.
and Mao, Zhiqiang
and Lau, Chun Ning
and Lu, Yuanming
and Trivedi, Nandini
and Vald{\'e}s Aguilar, Rolando
and Hight Walker, Angela R.},
title={Distinct magneto-{Raman} signatures of spin-flip phase transitions in {CrI$_{3}$}},
journal={Nat. Commun.},
year={2020},
month={Aug},
day={03},
volume={11},
number={1},
pages={3879},
issn={2041-1723},
doi={10.1038/s41467-020-17320-3},
url={https://doi.org/10.1038/s41467-020-17320-3}
}

@article{coh2023,
  title = {Classification of materials with phonon angular momentum and microscopic origin of angular momentum},
  author = {Coh, Sinisa},
  journal = {Phys. Rev. B},
  volume = {108},
  issue = {13},
  pages = {134307},
  numpages = {8},
  year = {2023},
  month = {Oct},
  publisher = {American Physical Society},
  doi = {10.1103/PhysRevB.108.134307},
  url = {https://doi.org/10.1103/PhysRevB.108.134307}
}

@article{schaack1976,
doi = {10.1088/0022-3719/9/11/009},
url = {https://doi.org/10.1088/0022-3719/9/11/009},
year = {1976},
month = {jun},
publisher = {},
volume = {9},
number = {11},
pages = {L297},
author = {G Schaack},
title = {Observation of circularly polarized phonon states in an external magnetic field},
journal = {J. Phys. C: Solid State Phys.}
}

@Article{zhang2023,
author={Zhang, Tiantian
and Huang, Zhiheng
and Pan, Zitian
and Du, Luojun
and Zhang, Guangyu
and Murakami, Shuichi},
title={Weyl Phonons in Chiral Crystals},
journal={Nano Lett.},
year={2023},
month={Aug},
day={23},
publisher={American Chemical Society},
volume={23},
number={16},
pages={7561},
issn={1530-6984},
doi={10.1021/acs.nanolett.3c02132},
url={https://doi.org/10.1021/acs.nanolett.3c02132}
}

@Article{ishito2023b,
author={Ishito, Kyosuke
and Mao, Huiling
and Kousaka, Yusuke
and Togawa, Yoshihiko
and Iwasaki, Satoshi
and Zhang, Tiantian
and Murakami, Shuichi
and Kishine, Jun-ichiro
and Satoh, Takuya},
title={Truly chiral phonons in {$\alpha$-HgS}},
journal={Nat. Phys.},
year={2023},
month={Jan},
day={01},
volume={19},
number={1},
pages={35},
issn={1745-2481},
doi={10.1038/s41567-022-01790-x},
url={https://doi.org/10.1038/s41567-022-01790-x}
}

@Article{spirito2024,
author ="Spirito, Davide and Marras, Sergio and Martín-García, Beatriz",
title  ="Lattice dynamics in chiral tellurium by linear and circularly polarized {Raman} spectroscopy: crystal orientation and handedness",
journal  ="J. Mater. Chem. C",
year  ="2024",
volume  ="12",
issue  ="7",
pages  ="2544",
publisher  ="The Royal Society of Chemistry",
doi  ="10.1039/D3TC04333A",
url  ="http://dx.doi.org/10.1039/D3TC04333A"
}

@article{kumar2020,
doi = {10.1088/1361-648X/ab9a7a},
url = {https://doi.org/10.1088/1361-648X/ab9a7a},
year = {2020},
month = {jul},
publisher = {IOP Publishing},
volume = {32},
number = {41},
pages = {415702},
author = {Kumar, Deepu and Singh, Birender and Kumar, Rahul and Kumar, Mahesh and Kumar, Pradeep},
title = {Anisotropic electron--photon--phonon coupling in layered {MoS$_{2}$}},
journal = {J. Phys.: Condens. Matter}
}

@article{ishioka2013,
  title = {Raman generation of coherent phonons of anatase and rutile {TiO$_{2}$} photoexcited at fundamental absorption edges},
  author = {Ishioka, Kunie and Petek, Hrvoje},
  journal = {Phys. Rev. B},
  volume = {86},
  issue = {20},
  pages = {205201},
  numpages = {6},
  year = {2012},
  month = {Nov},
  publisher = {American Physical Society},
  doi = {10.1103/PhysRevB.86.205201},
  url = {https://doi.org/10.1103/PhysRevB.86.205201}
}

@article{stevens2002,
  title = {Coherent phonon generation and the two stimulated {Raman} tensors},
  author = {Stevens, T. E. and Kuhl, J. and Merlin, R.},
  journal = {Phys. Rev. B},
  volume = {65},
  issue = {14},
  pages = {144304},
  numpages = {4},
  year = {2002},
  month = {Mar},
  publisher = {American Physical Society},
  doi = {10.1103/PhysRevB.65.144304},
  url = {https://doi.org/10.1103/PhysRevB.65.144304}
}

@article{han2023,
  title={Observing Axial Chirality of Chiral Single-Wall Carbon Nanotubes by Helicity-Dependent {Raman} Spectra},
  author={Han, Shiyi and Hung, Nguyen Tuan and Xie, Ying and Saito, Riichiro and Zhang, Jin and Tong, Lianming},
  journal={Nano Lett.},
  volume={23},
  number={18},
  pages={8454},
  year={2023},
  publisher={ACS Publications},
doi = {10.1021/acs.nanolett.3c01791},
  url = { https://doi.org/10.1021/acs.nanolett.3c01791}
}

@Article{yu2025,
author={Yu, Hao
and Li, Xinjie
and Bie, Ya-Qing
and Yan, Luo
and Zhou, Liujiang
and Yu, Peng
and Yang, Guowei},
title={Quantum metric third-order nonlinear Hall effect in a non-centrosymmetric ferromagnet},
journal={Nat. Commun.},
year={2025},
month={Aug},
day={18},
volume={16},
number={1},
pages={7698},
issn={2041-1723},
doi={10.1038/s41467-025-63133-7},
url={https://doi.org/10.1038/s41467-025-63133-7}
}

@article{lin.peng.2022,
doi = {10.1088/1674-1056/ac6eed},
url = {https://doi.org/10.1088/1674-1056/ac6eed},
year = {2022},
month = {aug},
publisher = {Chinese Physical Society and IOP Publishing Ltd},
volume = {31},
number = {8},
pages = {087506},
author = {Lin, Zhongchong and Peng, Yuxuan and Wu, Baochun and Wang, Changsheng and Luo, Zhaochu and Yang, Jinbo},
title = {Magnetic van der {Waals} materials: Synthesis, structure, magnetism, and their potential applications},
journal = {Chin. Phys. B}
}

@article{jiang.saito.07,
  title = {Exciton-photon, exciton-phonon matrix elements, and resonant {Raman} intensity of single-wall carbon nanotubes},
  author = {Jiang, J. and Saito, R. and Sato, K. and Park, J. S. and Samsonidze, Ge. G. and Jorio, A. and Dresselhaus, G. and Dresselhaus, M. S.},
  journal = {Phys. Rev. B},
  volume = {75},
  issue = {3},
  pages = {035405},
  numpages = {10},
  year = {2007},
  month = {Jan},
  publisher = {American Physical Society},
  doi = {10.1103/PhysRevB.75.035405},
  url = {https://link.aps.org/doi/10.1103/PhysRevB.75.035405}
}

@article{ZHANG2025c,
title = {Strategy for enhancing ferromagnetism in {2D} {Fe$_5$GaTe$_2$} and {Fe$_5$GeTe$_2$}},
journal = {Phys. Lett. A},
volume = {552},
pages = {130671},
year = {2025},
issn = {0375-9601},
doi = {10.1016/j.physleta.2025.130671},
url = {https://doi.org/10.1016/j.physleta.2025.130671},
author = {Nan Zhang and Zheng-Zhe Lin and Xi Chen},
}

@article{brennan2024,
  title={Strong Surface-Enhanced Coherent Phonon Generation in van der {Waals} Materials},
  author={Brennan, Christian
and Joly, Alan G.
and Wang, Chih-Feng
and Xie, Ti
and O'Callahan, Brian T.
and Crampton, Kevin
and Teklu, Alem
and Shi, Leilei
and Hu, Ming
and Zhang, Qian
and Kuthirummal, Narayanan
and Arachchige, Hasitha Suriya
and Chaturvedi, Apoorva
and Zhang, Hua
and Mandrus, David
and Gong, Cheng
and Gong, Yu},
  journal={J. Phys. Chem. Lett.},
  volume={15},
  number={42},
  pages={10442},
  year={2024},
doi={doi: 10.1021/acs.jpclett.4c02208
},
url={https://doi.org/10.1021/acs.jpclett.4c02208},
  publisher={ACS Publications}
}

@article{Bera2024,
  title = {Anisotropic magnetization dynamics in {Fe$_{5}$GeTe$_{2}$} at room temperature},
  author = {Bera, Alapan and Jana, Nirmalya and Agarwal, Amit and Mukhopadhyay, Soumik},
  journal = {Phys. Rev. B},
  volume = {110},
  issue = {22},
  pages = {224401},
  numpages = {15},
  year = {2024},
  month = {Dec},
  publisher = {American Physical Society},
  doi = {10.1103/PhysRevB.110.224401},
  url = {https://link.aps.org/doi/10.1103/PhysRevB.110.224401}
}

@article{Chakkar2024,
  title = {Broken weak and strong spin rotational symmetries and tunable interactions between phonons and the continuum in {Cr$_{2}$Ge$_{2}$Te$_{6}$}},
  author = {Chakkar, Atul G. and Kumar, Deepu and Kumar, Pradeep},
  journal = {Phys. Rev. B},
  volume = {109},
  issue = {13},
  pages = {134406},
  numpages = {11},
  year = {2024},
  month = {Apr},
  publisher = {American Physical Society},
  doi = {10.1103/PhysRevB.109.134406},
  url = {https://link.aps.org/doi/10.1103/PhysRevB.109.134406}
  
}

@article{Lan2025,
  title = {Thermomagnetic irreversibility in a {Cr$_{1.45}$Te$_{2}$} crystal: Role of spin-phonon coupling},
  author = {Lan, Ruihuan and Luo, Xuan and Zhou, Nan and Wang, Aile and Cheng, Ming and Liu, Lanxin and Pan, Yongqiang and Zhang, Ranran and Li, Jingxin and Hou, Yubin and Song, Wenhai and Lu, Qingyou and Sun, Yuping},
  journal = {Phys. Rev. B},
  volume = {112},
  issue = {10},
  pages = {104414},
  numpages = {14},
  year = {2025},
  month = {Sep},
  publisher = {American Physical Society},
  doi = {10.1103/qppq-qsx7},
  url = {https://link.aps.org/doi/10.1103/qppq-qsx7}
}

@article{Tiwari2024,
  title = {Phase transition in the two-dimensional {Heisenberg} ferromagnet {Fe$_{3}$GeTe$_{2}$} with long-range interaction},
  author = {Tiwari, Ankita and Ahn, Hyobin and Kumar, Birendra and Saini, Jyoti and Srivastava, Pawan Kumar and Singh, Budhi and Lee, Changgu and Ghosh, Subhasis},
  journal = {Phys. Rev. B},
  volume = {109},
  issue = {2},
  pages = {L020407},
  numpages = {6},
  year = {2024},
  month = {Jan},
  publisher = {American Physical Society},
  doi = {10.1103/PhysRevB.109.L020407},
  url = {https://link.aps.org/doi/10.1103/PhysRevB.109.L020407}
}

@article{joe.yang.19,
title = {First-principles study of ferromagnetic metal {Fe$_{5}$GeTe$_{2}$}},
journal = {Nano Mater. Sci.},
volume = {1},
number = {4},
pages = {299},
year = {2019},
issn = {2589-9651},
doi = {10.1016/j.nanoms.2019.09.009},
url = {https://doi.org/10.1016/j.nanoms.2019.09.009},
author = {Minwoong Joe and Unchun Yang and Changgu Lee}
}

@misc{nair.mallik.25,
Author = {Sreelakshmi M. Nair and Aabhaas Vineet Mallik and R. S. Patel},
Title = {Observation of {Raman} anomaly and characterization of magnetic phases in van der {Waals} ferromagnet {Fe$_5$GeTe$_2$}},
Year = {2025},
Eprint = {arXiv:2512.08665},
}

@article{McCreary2020b,
  title = {Quasi-two-dimensional magnon identification in antiferromagnetic {FePS$_{3}$} via magneto-{Raman} spectroscopy},
  author = {McCreary, Amber and Simpson, Jeffrey R. and Mai, Thuc T. and McMichael, Robert D. and Douglas, Jason E. and Butch, Nicholas and Dennis, Cindi and Vald\'es Aguilar, Rolando and Hight Walker, Angela R.},
  journal = {Phys. Rev. B},
  volume = {101},
  issue = {6},
  pages = {064416},
  numpages = {10},
  year = {2020},
  month = {Feb},
  publisher = {American Physical Society},
  doi = {10.1103/PhysRevB.101.064416},
  url = {https://link.aps.org/doi/10.1103/PhysRevB.101.064416}
}

@article{Liu2023,
  title = {Theoretical investigations on the magneto-{Raman} effect of {CrI$_{3}$}},
  author = {Liu, Shuang and Long, Meng-Qiu and Wang, Yun-Peng},
  journal = {Phys. Rev. B},
  volume = {108},
  issue = {18},
  pages = {184414},
  numpages = {6},
  year = {2023},
  month = {Nov},
  publisher = {American Physical Society},
  doi = {10.1103/PhysRevB.108.184414},
  url = {https://link.aps.org/doi/10.1103/PhysRevB.108.184414}
}

@article{kong2024,
  title={First-principles study of the magneto-{Raman} effect in van der {Waals} layered magnets},
  author={Kong, Xiangru and Ganesh, Panchapakesan and Liang, Liangbo},
  journal={npj 2D Mater. Appl.},
  volume={8},
  number={1},
  pages={82},
  year={2024},
  url={https://www.nature.com/articles/s41699-024-00515-3#citeas},
  doi={https://doi.org/10.1038/s41699-024-00515-3},
  publisher={Nature Publishing Group UK London}
}

@article{fernandez2019,
  title={Symmetry-breaking interlayer {Dzyaloshinskii--Moriya} interactions in synthetic antiferromagnets},
  author={Fern{\'a}ndez-Pacheco, Amalio and Vedmedenko, Elena and Ummelen, Fanny and Mansell, Rhodri and Petit, Doroth{\'e}e and Cowburn, Russell P},
  journal={Nat. Mater.},
  volume={18},
  number={7},
  pages={679},
  year={2019},
  doi={https://doi.org/10.1038/s41563-019-0386-4},
  url={https://www.nature.com/articles/s41563-019-0386-4#citeas},
  publisher={Nature Publishing Group UK London}
}

@article{nembach2015,
  title={Linear relation between {Heisenberg} exchange and interfacial {Dzyaloshinskii--Moriya} interaction in metal films},
  author={Nembach, Hans T and Shaw, Justin M and Weiler, Mathias and Ju{\'e}, Emilie and Silva, Thomas J},
  journal={Nat. Phys.},
  volume={11},
  number={10},
  pages={825},
  year={2015},
  doi={https://doi.org/10.1038/nphys3418},
  url={https://www.nature.com/articles/nphys3418#citeas},
  publisher={Nature Publishing Group UK London}
}

@article{zhang2022tuning,
  title={Tuning the exchange bias effect in 2D van der Waals ferro-antiferromagnetic {Fe$_3$GeTe$_2$/CrOCl} heterostructures},
  author={Zhang, Tianle and Zhang, Yujun and Huang, Mingyuan and Li, Bo and Sun, Yinghui and Qu, Zhe and Duan, Xidong and Jiang, Chengbao and Yang, Shengxue},
  journal={Adv. Sci.},
  volume={9},
  number={11},
  pages={2105483},
  year={2022},
  doi={ https://doi.org/10.1002/advs.202105483Digital Object Identifier (DOI)},
  url={https://advanced.onlinelibrary.wiley.com/doi/full/10.1002/advs.202105483},
  publisher={Wiley Online Library}
}


\clearpage
\newpage

\onecolumngrid

\begin{center}
  \textbf{\Large Supplemental Material}\\[.3cm]
  \textbf{\large Anisotropic electron--phonon coupling and chiral phonons \\[.2cm] in van der Waals room temperature ferromagnet Fe$_{5}$GeTe$_{2}$}\\[.3cm]
  Smrutiranjan~Mekap$^{1}$, Andrzej~Ptok$^{2}$, Jyoti~Saini$^{3}$, Changgu~Lee$^{4,5}$, Subhasis~Ghosh$^{3}$, and Anushree~Roy$^{1}$ \\[.2cm]
  {\itshape
${}^{1}$Department of Physics, Indian Institute of Technology Kharagpur, Kharagpur 721302, India \\[.1cm]
${}^{2}$Institute of Nuclear Physics, Polish Academy of Sciences, ul. W. E. Radzikowskiego 152, 31-342 Krak\'{o}w, Poland \\[.1cm]
${}^{3}$School of Physical Sciences, Jawaharlal Nehru University, New Delhi-110067, India\\[.1cm]
${}^{4}$SKKU Advanced Institute of Nanotechnology (SAINT), Sungkyunkwan University, Suwon 16419, Republic of Korea \\[.1cm]
${}^{5}$School of Mechanical Engineering, Sungkyunkwan University, Suwon 16419, Republic of Korea \\[.1cm]
  }
  (Dated: \today)
\\[0.3cm]
\end{center}

\setcounter{equation}{0}
\renewcommand{\theequation}{S\arabic{equation}}
\setcounter{figure}{0}
\renewcommand{\thefigure}{S\arabic{figure}}
\setcounter{section}{0}
\renewcommand{\thesection}{S\arabic{section}}
\setcounter{table}{0}
\renewcommand{\thetable}{S\arabic{table}}
\setcounter{page}{1}

\begin{figure}[!ht]
\centering
\includegraphics[width=0.35\linewidth]{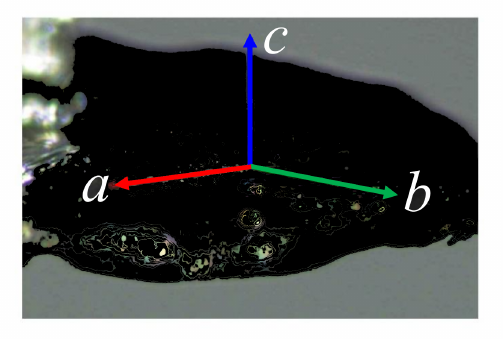}
\caption{Crystallographic axes orientation of the actual sample used for all Raman
analysis. 
}
\label{fig.orientation}
\end{figure}
\begin{figure}[!ht]
\centering
\includegraphics[width=0.7\linewidth]{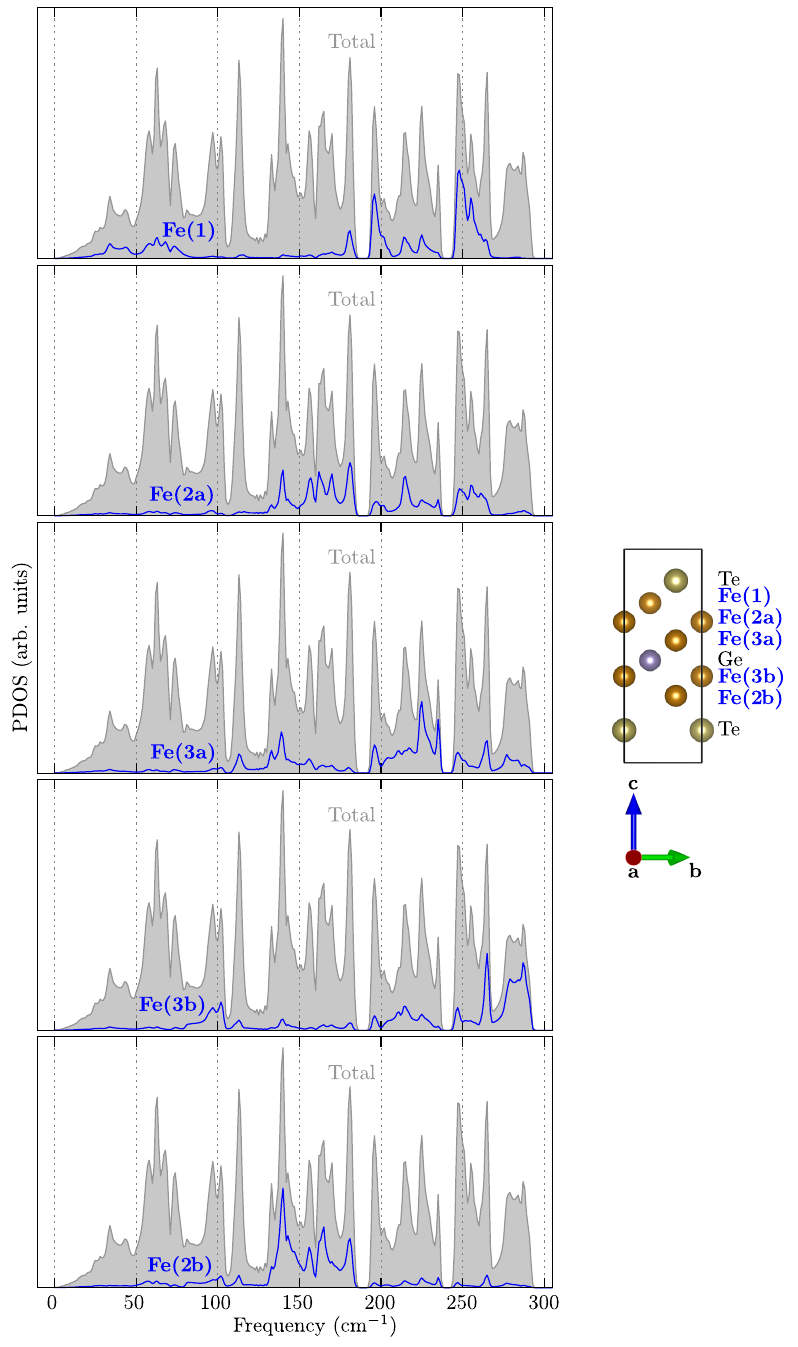}
\caption{
Partial phonon density of states for iron atoms, depend on their position within the crystal structure (right panel).
Results are obtained for F5GT model with P3m1 symmetry, within DFT+U calculations.
}
\label{fig.ph_fedos}
\end{figure}
\begin{figure}[!ht]
\centering
\includegraphics[width=0.8\linewidth]{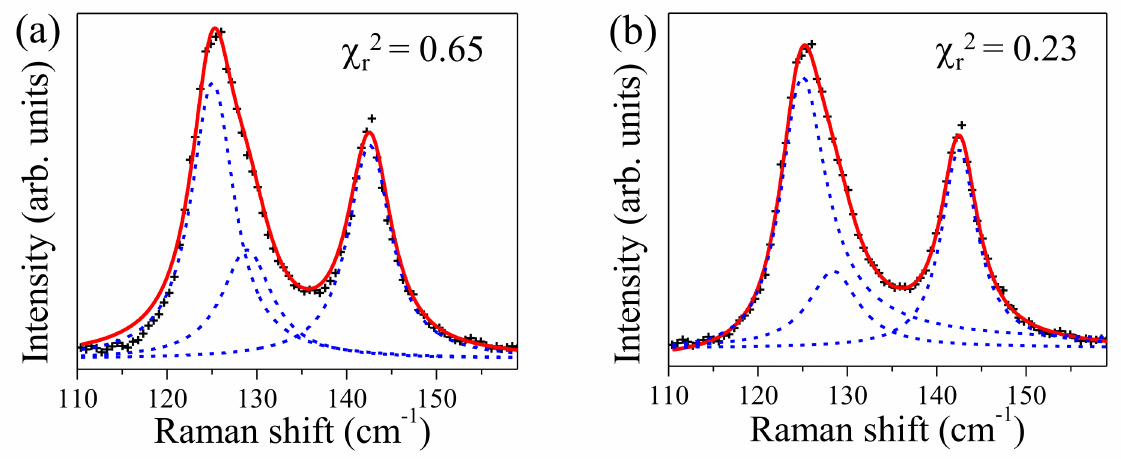}
\caption{
Magnified view of Raman spectrum at $80$~K between $110$ and $160$~cm$^{-1}$. 
In both panels ``+'' symbols are the spectral data, whereas the red curves are the best fits to the data points using (a) a Lorentzian function for each of A$_1$(1), E(2), and A$_1$(2) modes, and (b) Fano line shapes for both A$_1$ modes and Lorentizian functions for the E modes. 
In both panels, the deconvoluted components are shown by the blue dashed lines. 
The quality of the fit was evaluated using the reduced chi-square, $\chi^2_r$, values available in the insets.}
\label{fig.Fano_fit}
\end{figure}
\begin{figure}[!ht]
\centering
\includegraphics[width=0.7\linewidth]{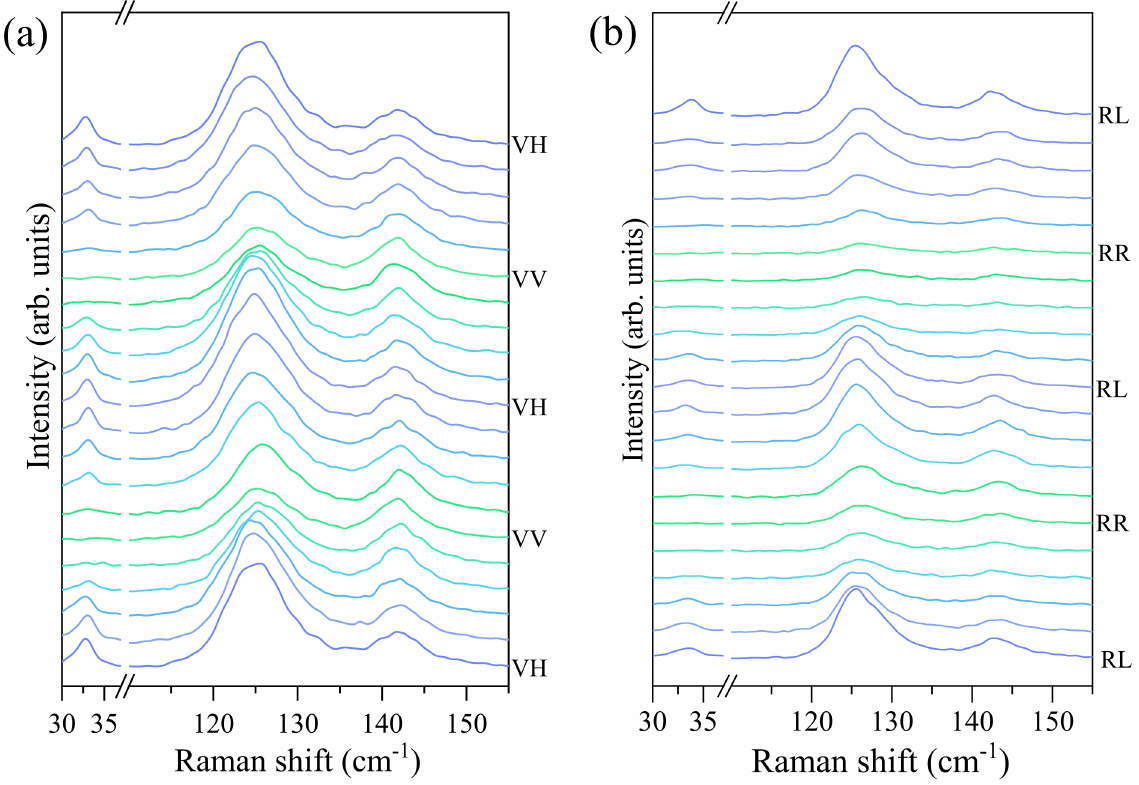}
\caption{
Full-angle polarized Raman spectra at $80$~K of (a) linear and (b) circular polarization.
\label{fig.full_angle_polar}
} 
\end{figure}
\begin{figure}[!ht]
\centering
\includegraphics[width=\linewidth]{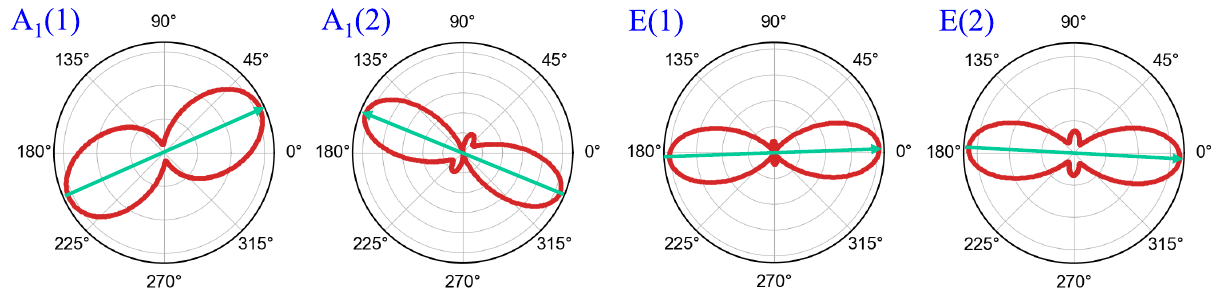}
\caption{
Theoretically obtained angle-dependent plots of the Raman intensity for modes correspond to the same presented in Fig.~\ref{fig.angle_raman} in the main text.
Results are obtained for single F5GT octuple monolayer, and laser energy $E_{ex} = 2.33$~eV.}
\label{fig.angle_raman_theo}
\end{figure}
\begin{figure}[!hb]
\centering
\includegraphics[width=0.55\linewidth]{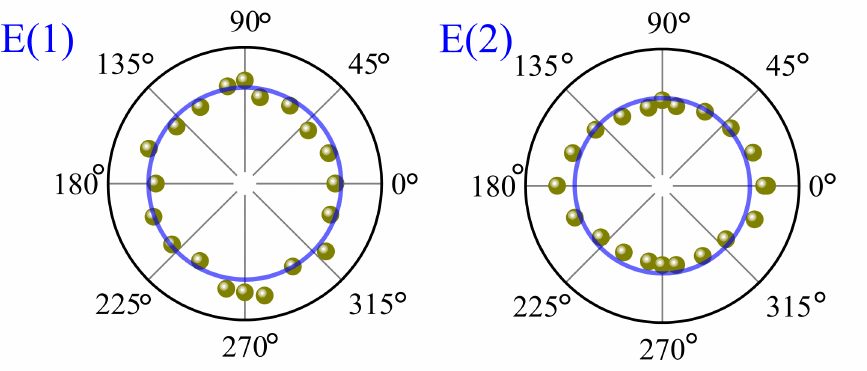}
\caption{
The angle-dependent plots of normalized Raman intensity for the discussed E modes (as labeled) at $80$~K, with $2.54$~eV laser excitation.
The data points are fitted with Eqs.~(\ref{eq.dephi}), and shown by blue curves.
\label{fig.angle_raman_488}}
\end{figure}
\begin{figure}[!ht]
\centering
\includegraphics[width=0.45\linewidth]{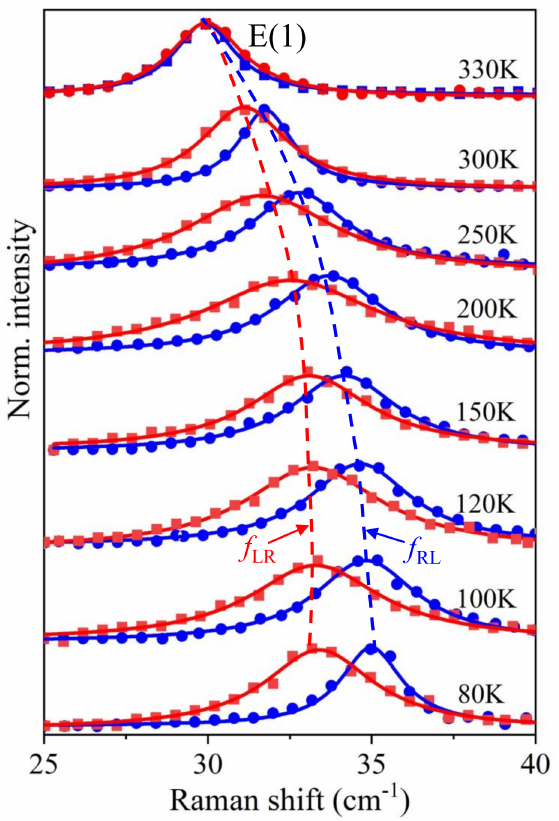}
\caption{
The helicity-resolved Raman spectra
of the E(1) mode for various temperatures. The blue and red dashed curves mark the shift in the spectra in RL and LR configurations with temperature.
\label{E mode_split_temp}}
\end{figure}
\begin{figure}[!ht]
\centering
\includegraphics[width=0.75\linewidth]{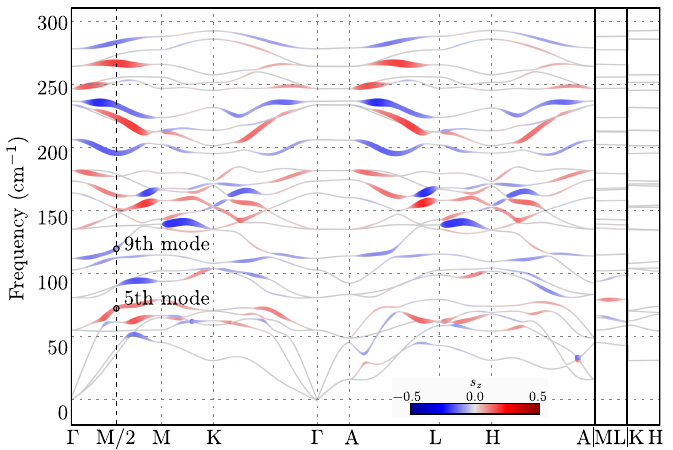}
\caption{
Theoretically calculated the phonon circular polarization along the Brillouin zone high symmetry directions.
Colored fat-bands correspond to the branches realized the chiral phonons.
Black points show phonon branches for wavevector M/2 (vertical black dashed line) analyzed in Fig.~\ref{fig.trace}.
\label{fig.full_circ}}
\end{figure}
\begin{figure}[!ht]
\centering
\includegraphics[width=\linewidth]{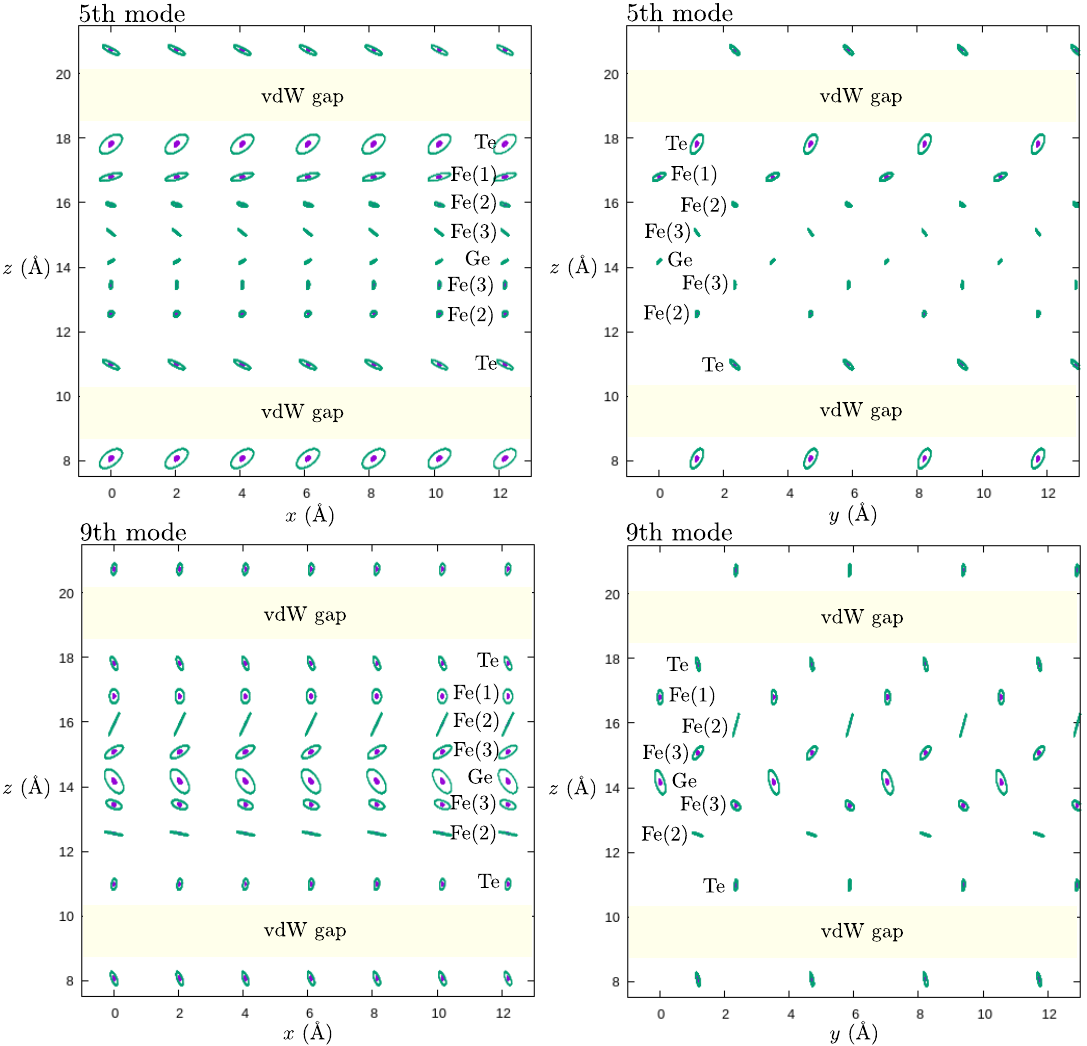}
\caption{
Example of the atomic trajectories defined by phonon from the 5th and 9th branches at wavevector equal M/2 (as labeled), marked by black circles in Fig.~\ref{fig.full_circ}.
Projection of the bulk system on $xz$ and $yz$ planes (left and right panel, respectively).
Equilibrium positions are given by purple points, while traces are shown by green lines.
As we can see, for some atoms, the trace is given by an ellipsoidal shape, which is evidence of the realization of chiral phonons in F5GT.
\label{fig.trace}}
\end{figure}

In this Supplemental Material, we present additional results:
\begin{itemize}
\item Sec.~\ref{sec.methods} -- Methods and techniques description.
\begin{itemize}
    \item Fig.~\ref{fig.orientation} Crystallographic axes orientation of the actual sample
    \item Fig.~\ref{fig.setup} Experimental setups.
\end{itemize}
\item Sec.~\ref{sec.system_para} -- Theoretically obtained system parameters.
\begin{itemize}
    \item Fig.~\ref{fig.model} F5GT crystal structure used in numerical calculations.
\end{itemize}
\item Sec.~\ref{sec.latt_dynam} -- Additional results related to the lattice dynamics:
\begin{itemize}
    \item Fig.~\ref{fig.irr_gamma} -- Schematic representation of vibrations corresponding to the Raman active modes.
    \item Fig.~\ref{fig.ph_fedos} --Partial phonon density of states for iron atoms.
\end{itemize}
\item Sec.~\ref{sec.raman_spec} -- Additional results from the Raman scattering measurements:
\begin{itemize}
\item Fig.~\ref{fig.cleaved} -- Raman spectra of cleaved sample.
    \item Fig.~\ref{fig.Fano_fit} -- Fano line shape.
    \item Tab.~\ref{tab.irr_gamma} -- Comparison of the experimental and theoretical characteristic frequencies at $\Gamma$.
\end{itemize}
\item Sec.~\ref{sec.raman_intens} --  Raman intensities
\begin{itemize}
    \item Sec.~\ref{sec.full_angle}: Full angle resolved Raman spectra
\item
Fig.~\ref{fig.full_angle_polar}--Characteristic spectra recorded at various linear and circular polarization configurations.
    \item Sec.~\ref{sec.Raman_tensors}: Calculated polarization-dependent Raman intensity.
    \item Fig.~\ref{fig.angle_raman_theo} -- Theoretically obtained angle-dependent plots of the Raman intensity.
    \item Tab.~\ref{tab.weight_factors} -- Fitted parameters at various temperatures.
\end{itemize}
\item Sec.~\ref{sec.fano_param} -- Study of the Fano parameter for asymmetric peaks.
\begin{itemize}
    
    \item Fig.~\ref{fig.fano_asymmetry_energy} -- Evolution of Fano parameters with laser excitation energy.
    \item Fig.~\ref{fig.angle_raman_488} --Polar plot of normalized Raman intensity with 2.54  eV exciatation energy at 80 K.
\end{itemize}
\item Sec.~\ref{sec.chiral_phon} -- Additional results related to the chiral phonons:
\begin{itemize}
    \item Fig.~\ref{E mode_split_temp} -- Evolution of $\Delta f$ with temperature.
    \item Fig.~\ref{fig.full_circ} -- The phonon circular polarization.
    \item Fig.~\ref{fig.trace} -- The atomic trajectories realized by atomic vibrations associated with the chiral phonon.
\end{itemize}
\end{itemize}


\newpage

\section{Methods and techniques}
\label{sec.methods}

\subsection{Experimental details}

\vspace{1cm} 

\vspace{3cm}

\begin{figure}[!h]
\centering
\includegraphics[width=\linewidth]{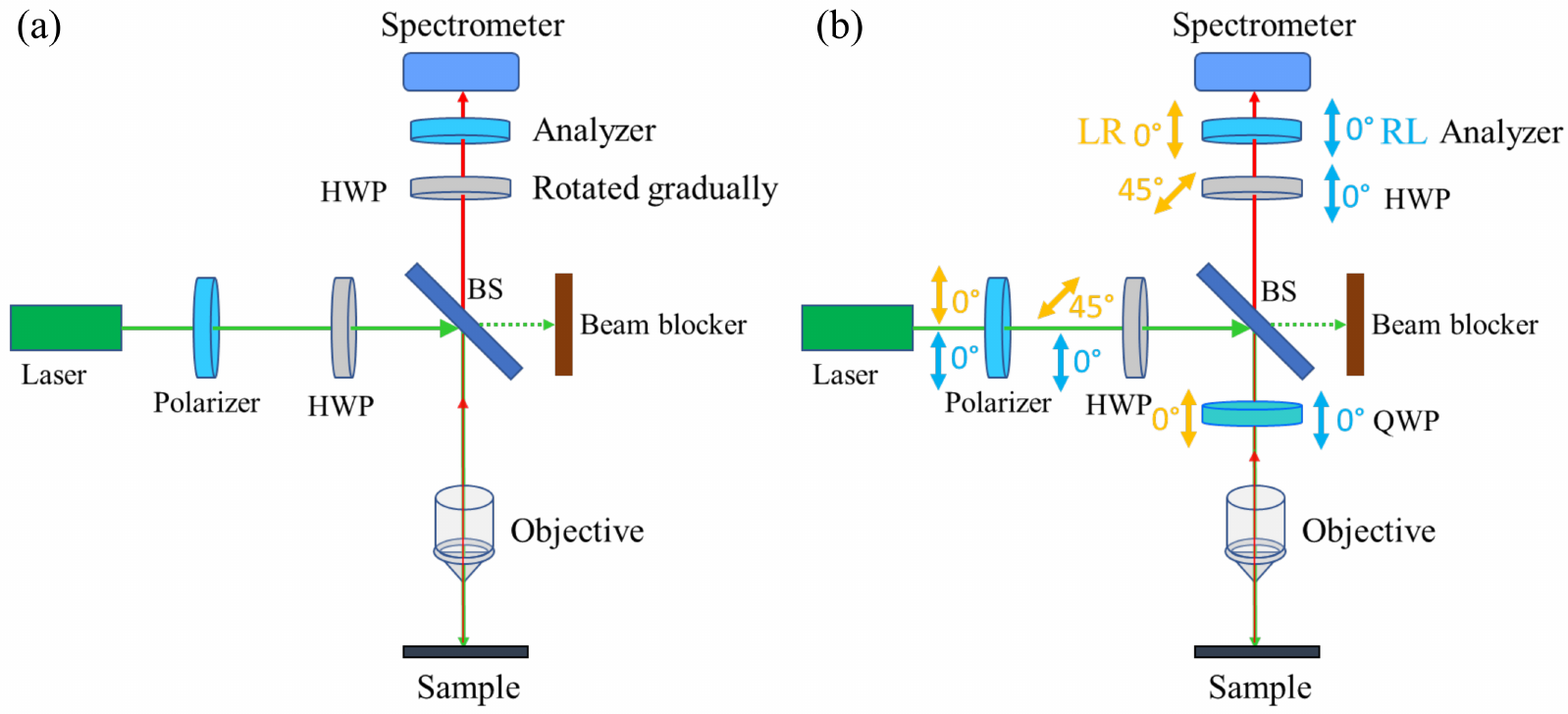}
\caption{
Experimental setups for (a) linear and (b) circular polarization dependent Raman measurements. HWP: half wave plate, QWP: quarter wave plate, BS: beam splitter.} 
\label{fig.setup}
\end{figure}

\newpage

\section{System parameter}
\label{sec.system_para}

\begin{figure}[!ht]
\centering
\includegraphics[width=0.5\linewidth]{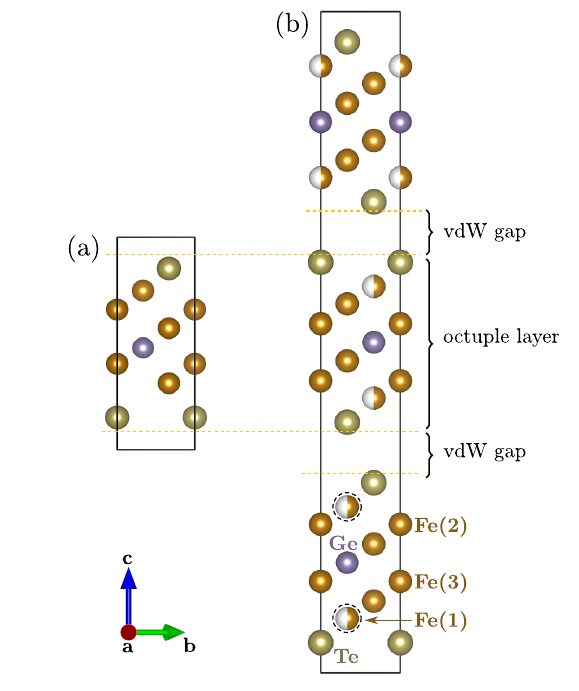}
\caption{
Panel (a) shows the model of the F5GT crystal structure used in numerical calculations. 
Such a bulk model with P3m1 symmetry is based on the single octuple layer in structure with R$\bar{3}$m symmetry, presented on panel (b).
Such a structure can be formed from the original one, when the Fe(1) atom fills only one (top or bottom) position.
}
\label{fig.model}
\end{figure}

Due to the theoretical limitation, we investigate F5GT using a bulk model based on a single octuple layer periodically repeated along the $c$ direction.
This is primarily associated with the real crystal structure with structural disorder (discussed in the main text).
The study of lattice dynamics is technically impossible for the system with fractional occupancy.
The proposed model of the crystal with the trigonal P3m1 symmetry (space group No.~156) corresponds to the structure with AA stacking (see Fig.~\ref{fig.model}).

Experimental lattice constants of F5GT were reported as $a = 4.04$~\AA, and $c = 29.19$~\AA~\cite{stahl2018}.
After optimization of a single octuple layer, i.e., for the simplified model, the lattice constants were obtained as $a_\text{S} = 3.963$~\AA\ ($4.062$~\AA), and $c_\text{S} =9.619$~\AA\ ($9.747$~\AA) within DFT (DFT+U) calculations.
As we can see, the theoretically obtained lattice constants of the simplified model excellently reproduce the experimental lattice of the ``full'' F5GT cell (i.e., $a_\text{S} \simeq a$ and $3 \; c_\text{S} \simeq c$).

\subsection*{Magnetic features}

The theoretical investigation indicates important role of the Fe$^{2+}$/Fe$^{3+}$ ratio~\cite{liu2022}.
For F5GT, the non-equivalent Fe positions affect the observed effective magnetic moments.
For example, in F3GT, there are two types of the $3d$ states: mostly localized states ($d_{z^{2}}$) and itinerant ($d_{xz/yz}$ and $d_{xy/x^{2}-y^{2}}$)~\cite{xu2020b,liu2022}.
According to the orbital occupation behavior of the localized states and the different coordination environments, two valence states of Fe atoms are realized: trivalent Fe$^{3+}$ and divalent Fe$^{2+}$.
Therefore, the resulting calculated net magnetic moments are $3.0$~$\mu_{B}$ and $2.6$~$\mu_{B}$ for Fe$^{3+}$ and Fe$^{2+}$, respectively~\cite{zhu2016,liu2022}.
F5GT exhibits a similar effect with three different magnetic moments, which depend on the positions of the Fe atoms (see Fig.~\ref{fig.model}).
In our DFT+U calculations, we obtain similar values of Fe magnetic moments: $\sim 2.53$~$\mu_{B}$ for Fe(1) and Fe(2), and $\sim 3$~$\mu_{B}$ for Fe(3). 
This result is comparable with previous theoretical studies~\cite{may2022,may2019a,ershadrad2022}.

\newpage

\section{Lattice dynamics}
\label{sec.latt_dynam}

\begin{figure}[!ht]
\centering
\includegraphics[width=\linewidth]{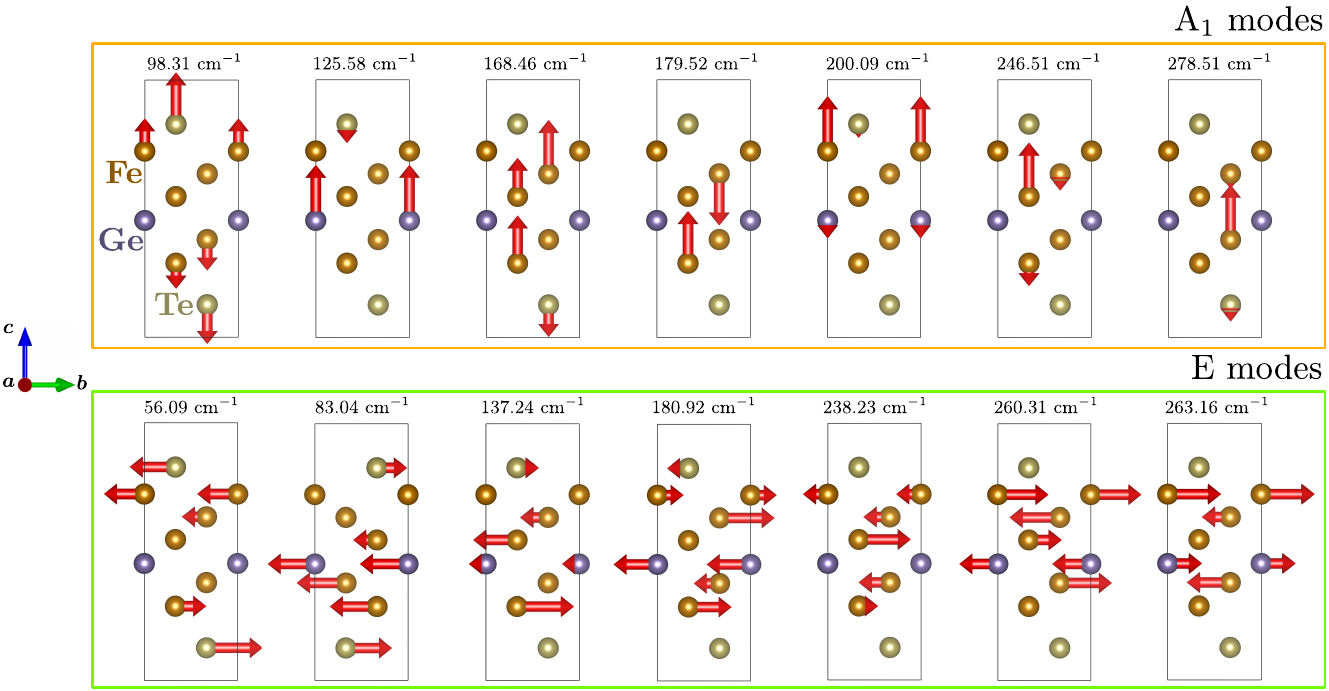}
\caption{
Raman active modes with A$_{1}$ and E symmetry (as labeled).
Results are obtained for F5GT model with P3m1 symmetry, within DFT+U calculations. 
}
\label{fig.irr_gamma}
\end{figure}

\newpage

\section{Raman spectroscopy}
\label{sec.raman_spec}

\subsection{Raman spectra of cleaved sample}
\label{cleave}
\begin{figure}[h]
\centering
\includegraphics[width=0.4\linewidth]{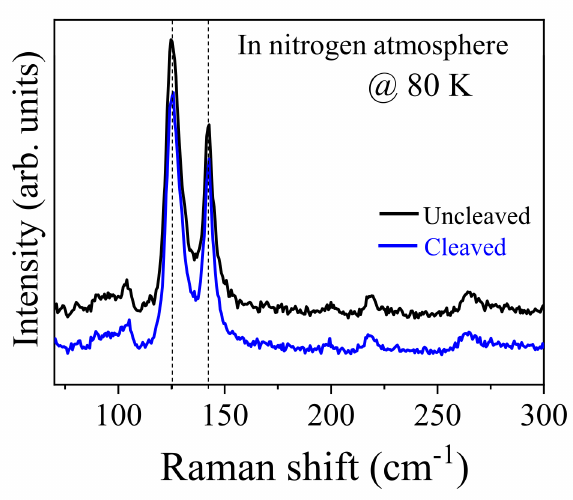}
	\caption{ Raman spectra recorded for F5GT of freshly cleaved sample (blue) and that recorded at a much later stage of full sets of experiments (black). Both spectra were recorded in a nitrogen atmosphere at $80$~K.}
	\label{fig.cleaved}
\end{figure}

The role of the naturally formed oxide layer on FGT systems in determining the  Raman spectrum of the compound has been debated in the literature \cite{zhang2025b,zhang2022tuning}. To address this issue, we compare the Raman spectrum of a freshly cleaved (within the glove box) F5GT sample with that recorded at a much later stage in Fig. \ref{fig.cleaved}. 
Both these spectra were collected at 80 K in a nitrogen atmosphere. The profiles of these two spectra are very similar and reproducible across multiple measurements over the duration of our experiment, indicating that the observed phonon features are intrinsic and not significantly altered by ambient exposure during our experiment.

\subsection{Fano line-shape}
\label{Fano_fit}

\newpage

\subsection{Calculated and experimentally measured Raman frequencies}

\begin{table}[!hb]
\caption{
Characteristic frequencies (cm$^{-1}$) and symmetries of irreducible representations (Irr) of the phonon modes at $\Gamma$ point for simplified P3m1 symmetry.
Experimental frequencies are obtained for $80$ and $300$~K, by fitting the experimental data as described in the main text.
\label{tab.irr_gamma}
}
\begin{ruledtabular}
\begin{tabular}{cccccc}
\multicolumn{2}{c}{DFT} & \multicolumn{2}{c}{DFT+U} & \multicolumn{2}{c}{Experimental} \\
Frequency & Irr & Frequency & Irr & $T = 80$~K & $T = 300$~K \\
\hline
56.87 & E & 56.09 & E & 33.6 & 31.25 \\
103.52 & E & 83.04 & E &93.6 & 92.2 \\
111.14 & A$_{1}$ & 98.31 & A$_{1}$ &104.1 & 99.1 \\
138.67 & A$_{1}$ & 125.58 & A$_{1}$ &124.8 & 122.5 \\
167.17 & E & 137.24 & E & 128.9& 126.0 \\
177.41 & A$_{1}$ & 168.46 & A$_{1}$ &142.9 & 140.7 \\
197.57 & E & 179.52 & A$_{1}$ & --- & --- \\
203.38 & A$_{1}$ & 180.92 & E & --- & --- \\
211.57 & A$_{1}$ & 200.09 & A$_{1}$ & 200.9 & 197.8 \\
236.03 & A$_{1}$ & 238.23 & E & 219.1 & 218.9 \\
254.41 & E & 246.51 & A$_{1}$ & 230.6 & 224.1 \\
274.88 & E & 260.31 & E & --- & --- \\
298.94 & A$_{1}$ & 263.16 & E &265.1 & 262.5 \\
310.04 & E & 278.51 & A$_{1}$ & 283.6& 281.4
\end{tabular}
\end{ruledtabular}
\end{table}

\section{Raman intensities for linear and circular polarization}
\label{sec.raman_intens}

\subsection{Full angle resolved Raman spectra}
\label{sec.full_angle}

\subsection{Calculated intensities}
\label{sec.Raman_tensors}

We begin by defining the crystal basis as
\begin{eqnarray}
X = ( 1 \; 0 \; 0 )^{T} \quad \text{and} \quad Y = ( 0 \; 1 \; 0 )^{T} .
\end{eqnarray}
For linearly polarized light, the angle-dependent intensity (for $\hat{e}_{i} = (0 \;1 \;0)^{T}$ and $\hat{e}_{s}^{T} =(\cos\varphi \;\sin\varphi \;\;0)$ with Raman tensors defined as in Eq.~(\ref{eq.raman_tensor}) of the main text, can be found as:
\begin{eqnarray}
I^{L} ( \text{A}_{1} , \varphi ) = \vert a \sin \varphi \vert^{2}, \quad \text{and} \quad I^{L} ( \text{E} ) = \vert c \vert^{2} .
\label{normal_tensor}
\end{eqnarray}
where the analyzer is rotated by an angle $\varphi$ measured from the X axis for different directions of scattered polarization in the sample plane. In Tab.~\ref{tab.ramanselect} we present the selection rules for the investigated modes.

A similar relation can be found for circularly polarized light, with right- and left-handed circular polarization $\sigma^{\pm} = \frac{1}{\sqrt{2}} ( 1 \; \pm i \; 0 )^{T}$.
The angular-dependent intensity (for $\hat{e}_{i} = \sigma^{+}$ and $\hat{e}_{s} = \sigma^{+} \cos \varphi + \sigma^{-} \sin \varphi$) is given as:
\begin{eqnarray}
I^{C} ( \text{A}_{1} , \varphi ) = \vert a \sin \phi \vert^{2}, \quad \text{and} \quad I^{C} ( \text{E} , \varphi ) = 2 \vert c \vert^{2} ,
\end{eqnarray}

For magnetic crystals, the $3m$ magnetic point group of P3m1 space group, the antisymmetrical imaginary elements are allowed in the Raman tensor as~\cite{cracknell1969}:
\begin{eqnarray}
DA_{1} = \left( \begin{array}{ccc}
    a & if & 0 \\
     -if & a & 0 \\
     0 & 0 & b \\
    \end{array}\right), \quad \text{and} \quad 
DE_{1} = \left( \begin{array}{ccc}
    d & id & ic \\
     id & -d & c \\
     ig & g & 0 \\
    \end{array}\right) 
    DE_{2} = \left( \begin{array}{ccc}
    e & -ie & -ij \\
     -ie & -e & j \\
     -ih & h & 0 \\
    \end{array}\right) .
\end{eqnarray}
For linearly polarized light, the angle dependent intensity can be found as
\begin{eqnarray}
\label{eq.da1}
I^{L} ( D\text{A}_{1} , \varphi ) = \vert a \sin \varphi \vert^{2}+ \vert f \cos \varphi \vert^{2}
\end{eqnarray}
and
\begin{eqnarray}
I^{L} (D\text{E}_{1},\varphi ) = \vert d\vert^{2}, \quad \text{and} \quad I^{L} (D\text{E}_{2},\varphi ) =\vert e\vert^{2}.
\end{eqnarray}

For the incoherent superposition of two degenerate E modes\\
\begin{eqnarray}
\label{eq.dephi}
I^{L}(D\text{E},\varphi)= \vert I^{L} (D\text{E}_{1},\varphi ) \vert+\vert I^{L}(D\text{E}_{2},\varphi)\vert= \vert d \vert^{2} +\vert e \vert^{2},
\end{eqnarray}
which is again polarization angle independent.

Several articles in the literature discuss the Raman generation of coherent phonons in the context of ultrafast phonon dynamics~\cite{ishioka2013,stevens2002}, especially under resonance conditions; however, such a physical concept of coherent superposition of tensors is rare in the context of spontaneous Raman scattering. Assuming such a coherent superposition of two degenerate E modes:
\begin{eqnarray}
\label{eq.dephi2} I^{L}(D\text{E},\varphi)= \vert u.(D\text{E}_{1})+v.(D\text{E}_{2} ) \vert^{2} =\vert u.D\vert^{2}+\vert v.E\vert^{2}+2\vert u.D\vert \vert v.E\vert\sin^{2} \varphi-2\vert u.D\vert \vert v.E \vert \cos^{2}\varphi.
\end{eqnarray}

The electron--phonon coupling can introduce additional phase factors to the tensor elements~\cite{pimenta2021,han2023}

\begin{eqnarray}
\nonumber DA_1^{mod} = \left( \begin{array}{ccc}
    a \exp \left( {i\alpha_{a}} \right) & if & 0 \\
     -if & a \exp \left( {i\alpha_{a^{}}} \right) & 0 \\
     0 & 0 & b \exp \left( {i\alpha_{b}} \right) \\
   \end{array}\right) , \quad \text{and} \quad 
DE^{mod}_{1} = \left( \begin{array}{ccc}
    d \exp \left( {i\alpha_{d}} \right) & id & ic \\
     id & -d \exp \left( {i\alpha_{d}} \right) & c \\
     ig & g & 0 \\
   \end{array}\right)\\
   \quad \text{and} \quad 
   DE^{mod}_{2} = \left(
   \begin{array}{ccc}
    e \exp \left( {i\alpha_{e}} \right) & -ie & -ij \\
     -ie & -e\exp \left( {i\alpha_{e}} \right) & j \\
     -ih & h & 0 \\
   \end{array}\right) .
\end{eqnarray}

\begin{eqnarray}
\label{eq.da1phi} I^{L} ( DA_1^{mod} , \varphi ) = \vert a \sin \varphi*\cos\alpha_{a}\vert^{2}+ \vert a\sin \varphi* \sin \alpha_{a}+f \cos \varphi \vert^{2}, \quad \text{and} \quad\\ I^{L} (D\text{E}_{1}^{mod},\varphi ) = \vert d \cos \alpha_{d}*\sin \varphi\vert^{2}+\vert d \sin \alpha_{d}*\sin \phi +d \cos \varphi\vert^{2}, \quad \text{and}\quad\\ I^{L} (D\text{E}_{2}^{mod},\varphi ) = \vert e \cos \alpha_{e}*\sin \varphi\vert^{2}+\vert e \sin \alpha_{e}*\sin \varphi -e \cos \varphi\vert^{2}.
\end{eqnarray}

For circularly polarized light, the angle-dependent intensity can be found as
\begin{eqnarray}
I^{C} ( D\text{A}_{1} , \varphi ) = \vert (a-f) \sin \varphi \vert^{2}, \quad \text{and} \quad\\ I^{C} (D\text{E}_{1},\varphi ) = 0, \quad \text{and} \quad I^{C} (D\text{E}_{2},\varphi ) =\vert 2e \cos \varphi \vert^{2}.
\end{eqnarray}

\vspace{4cm}

\begin{table}[!hb]
\caption{
\label{tab.ramanselect}
Selection rules for Raman-active modes.
}
\begin{ruledtabular}
\begin{tabular}{lcc}
configuration & A$_{1}$ & E \\
\hline 
$e_{x}$ in $e_{x}$ out (linear $\parallel$) & $|a|^{2}$ & $|c|^{2}$ \\
$e_{x}$ in $e_{y}$ out (linear $\perp$) & $0$ & $|c|^{2}$ \\
$\sigma^{+}$ in $\sigma^{+}$ out (cocircular) & $|a|^{2}$ & $0$ \\
$\sigma^{+}$ in $\sigma^{-}$ out (cross-circular) & $0$ & $2|c|^{2}$
\end{tabular}
\end{ruledtabular}
\end{table}

\newpage

\subsection{Polar dependence}
\label{sec.polar_dep}

\begin{table}[!hb]
\caption{
\label{tab.weight_factors}
Fitted parameter at various temperature using Eqn.~\ref{eq.da1} and~\ref{eq.dephi2} for Fig.~\ref{fig.angle_raman} in the main text.}
\begin{ruledtabular}
\begin{tabular}{clccc}
Raman & & 80 K & 300 K & 330 K\\
mode & & & \\
\hline
\multirow{4}{1em}{E(1)} & $d$ & $0.70 \pm 0.10 $ & $0.70 \pm 0.10$ & $0.6 \pm 0.2$ \\
& $e$    &   $0.40 \pm 0.02$ & $0.40 \pm 0.03$ & $0.64 \pm 0.15$\\
& $u$    &   $1.04 \pm 0.02$ & $1.14 \pm 0.02$ & $-$ \\
& $v$    &   $-0.34 \pm 0.15$ & $-0.41 \pm 0.02$& $-$ \\
\hline
\multirow{4}{1em}{E(2)} & $d$ & $0.70 \pm 0.10 $ & $0.70 \pm 0.10$& $0.64 \pm 0.14$ \\
& $e$    &   $0.40 \pm 0.01$ & $0.4 \pm 0.02$& $0.64 \pm 0.12$ \\
& $u$    &   $0.97 \pm 0.02$ & $1.00 \pm 0.03$ & $-$\\
& $v$    &   $-0.44 \pm 0.04$ & $-0.56 \pm 0.06$ & $-$ \\
\hline
\multirow{3}{1em}{A$_1$(1)} & $a$ & $0.72 \pm 0.02 $ & $0.75 \pm 0.01 $ & $0.93 \pm 0.53$\\
 & $f$ & $0.94 \pm 0.01$ & $0.95 \pm 0.02$ & $0.74 \pm 0.32$\\
& $\alpha_{a}$ (rad) & $-5.87\pm 1.00$ & $4.24\pm1.12$ & $1.74 \pm 0.92$\\
\hline
\multirow{3}{1em}{A$_1$(2)} & $a$ & $0.92 \pm 0.01$ & $0.88 \pm 0.01 $& $0.96 \pm 0.29$ \\
& $f$ &$0.93 \pm 0.01$ &$1.10 \pm 0.01$& $0.68 \pm 0.12$\\
& $\alpha_{a}$ (rad) & $2.85 \pm 1.00$ & $0.22\pm0.10$ & $-0.10 \pm 0.09$
\end{tabular}
\end{ruledtabular}
\end{table}

\newpage

\section{Fano parameter}
\label{sec.fano_param}

\subsection{Evolution of Fano parameters with laser excitation energies}

\begin{figure}[!ht]
\centering
\includegraphics[width=0.9\linewidth]{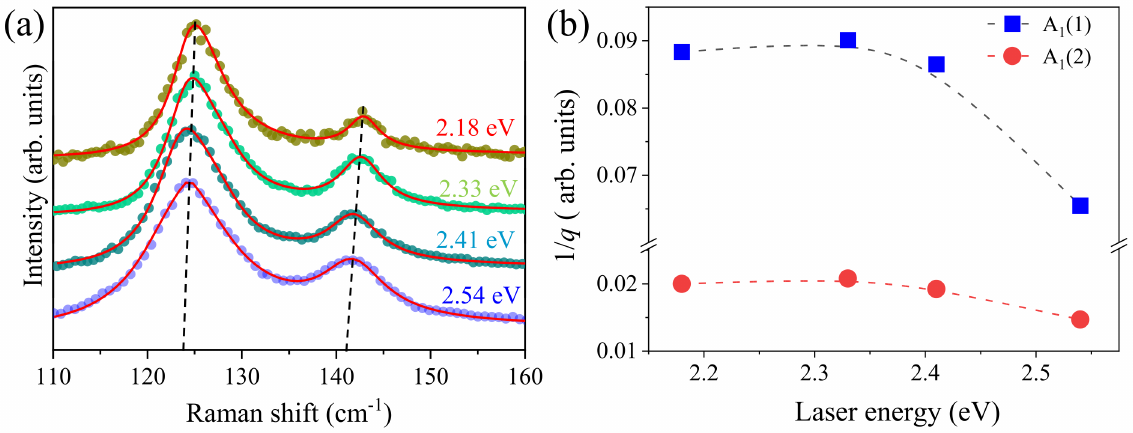}
\caption{(a) Magnified view of Raman spectrum at $80$~K between spectral range $110$ and $160$~cm$^{-1}$ with laser excitation energies. 
Symbols are the spectral data, whereas the red curves are the best fits to the data points using a Fano line shape for both A$_1$ modes and a Lorentzian function for the E(2) mode.
(b) Variation of the asymmetric parameter with laser excitation energy.
\label{fig.fano_asymmetry_energy}}
\end{figure}

Figure~\ref{fig.fano_asymmetry_energy} (a) plots Raman spectra of F5GT over the spectral range between $110$ and $160$~cm$^{-1}$ for excitation energies of the laser source at $2.18$, $2.33$, $2.41$, and $2.54$~eV. 
We observe a clear shift of both A$_1$ modes towards higher wavenumbers with the change in laser energy. 
The increase in the Fano asymmetry parameter $1/q$ results in a blue shift in the Raman mode, along with an asymmetry in the spectral profile in the same direction. 
The variations of this parameter with the laser excitation energy for the A$_1$(1) and A$_1$(2) modes are plotted in Fig.~\ref{fig.fano_asymmetry_energy} (b). 
It is evident that the value of the same parameter drops significantly for the laser energy $2.54$~eV.
Notably, the parameter exhibits a pronounced decrease at $2.54$~eV, indicating that electron--phonon coupling is significantly stronger for the excitation energies $2.18$~eV and $2.33$~eV and is suppressed at higher excitation energies.

\newpage

\section{Chiral phonons}
\label{sec.chiral_phon}

\vspace{2cm}

\end{document}